\documentclass[
reprint,
superscriptaddress,
nofootinbib,
amsmath, amssymb,
prd,
twocolumn,
]{revtex4-1}

\usepackage[utf8]{inputenc}
\usepackage{amsmath}
\usepackage{mathtools}
\usepackage{amssymb,amstext,amsfonts}
\usepackage{graphicx}
\usepackage{hyperref} %To link references from text
\usepackage{float}
\usepackage{microtype}
\usepackage{subfigure}
\usepackage{multirow}
\usepackage{lineno}
\usepackage{color}
\usepackage{comment}
\usepackage[normalem]{ulem}
\usepackage{lipsum}
\usepackage{bm}

\newcommand{\be}{ \begin{equation} }
\newcommand{\ee}{ \end{equation}}

\begin{document}

\title{Constraints on Kerr-Newman black holes from\\ merger-ringdown gravitational-wave observations}

\author{Gregorio Carullo}
\affiliation{Dipartimento di Fisica ``Enrico Fermi'', Universit\`a di Pisa, Pisa I-56127, Italy}
\affiliation{INFN sezione di Pisa, Pisa I-56127, Italy}

\author{Danny Laghi}
\affiliation{Dipartimento di Fisica ``Enrico Fermi'', Universit\`a di Pisa, Pisa I-56127, Italy}
\affiliation{INFN sezione di Pisa, Pisa I-56127, Italy}
\affiliation{Laboratoire des 2 Infinis - Toulouse (L2IT-IN2P3), Universit\'e de Toulouse, CNRS, UPS, F-31062 Toulouse Cedex 9, France}

\author{Nathan~K.~Johnson-McDaniel}
\affiliation{DAMTP, Centre for Mathematical Sciences,
    University of Cambridge, Wilberforce Road, Cambridge CB3 0WA, United Kingdom}
\affiliation{Department of Physics and Astronomy, University of Mississippi, University, Mississippi 38677, USA}

\author{Walter Del Pozzo}
\affiliation{Dipartimento di Fisica ``Enrico Fermi'', Universit\`a di Pisa, Pisa I-56127, Italy}
\affiliation{INFN sezione di Pisa, Pisa I-56127, Italy}

% This solution does not work because it prints '(and)' with unwanted parentheses.
%\collaboration{and}

% This solution does not work because it prints a comma before the second block
% \author{\\ \vspace{0.3cm} and \\}\noaffiliation

% This solution is incompatible with revtex4
%\let\c@affil\relax
%\makeatother
%\usepackage{authblk} % to have block-separated authors

\author{\'Oscar J.~C.~Dias}
\affiliation{STAG research centre and Mathematical Sciences, University of Southampton, UK}

\author{Mahdi Godazgar}
\affiliation{School of Mathematical Sciences, Queen Mary University of London, Mile End Road, London E1 4NS, UK.}

\author{Jorge E.~Santos}
\affiliation{DAMTP, Centre for Mathematical Sciences,
    University of Cambridge, Wilberforce Road, Cambridge CB3 0WA, United Kingdom}

\date{\today}

\begin{abstract}

We construct a template to model the post-merger phase of a binary black hole coalescence in the presence of a remnant $U(1)$ charge. 
We include the quasi-normal modes typically dominant during a binary black hole coalescence, $(\ell,m,n) = \{(2,2,0), (2,2,1)\}$ and also present analytical fits for the quasinormal mode frequencies of a Kerr-Newman black hole in terms of its spin and charge, here also including the $(3,3,0)$ mode.
Aside from astrophysical electric charge, our template can accommodate extensions of the Standard Model, such as a dark photon.
Applying the model to LIGO-Virgo detections, we find that we are unable to distinguish between the charged and uncharged hypotheses from a purely post-merger analysis of the current events. 
However, restricting the mass and spin to values compatible with the analysis of the full signal, we obtain a 90th percentile bound $\bar{q} < 0.33$ on the black hole charge-to-mass ratio, for the most favorable case of GW150914. 
Under similar assumptions, by simulating a typical loud signal observed by the LIGO-Virgo network at its design sensitivity, we assess that this model can provide a robust measurement of the charge-to-mass ratio only for values $\bar{q} \gtrsim 0.5$; here we also assume that the mode amplitudes are similar to the uncharged case in creating our simulated signal.
Lower values, down to $\bar{q} \sim 0.3$, could instead be detected when evaluating the consistency of the pre-merger and post-merger emission.

\end{abstract}

\maketitle

\section{Introduction}
The most generic family of regular, stationary, asymptotically flat, electrovacuum solutions in Einstein-Maxwell theory is the Kerr-Newman (KN) family of black holes (BHs)~\cite{1965JMP.....6..918N, Mazur_1982}. 
This solution extends the Kerr metric~\cite{Kerr:1963ud} and is uniquely characterised by the mass $M$, the dimensionless spin $\chi$, and charge-to-mass ratio $\bar{q}$, typically identified with the electric charge-to-mass ratio of the BH.
Astrophysical BHs are expected to carry negligible electric charge~\cite{Gibbons, Znajek, Palenzuela:2011es}. Although a rotating BH embedded in a magnetic field can selectively accrete electric charge, the maximum amount accreted through this effect is negligible for astrophysical values of magnetic fields~\cite{PhysRevD.10.1680}. 
Additionally, mechanisms such as vacuum polarization, breakdown pair production, and neutralisation from surrounding material prevent a stellar-mass BH from sustaining a large amount of electric charge~\cite{Znajek, Gibbons}.
Even if a significant amount of charge is acquired, it is dissipated on a time scale much shorter than the one probed by gravitational-wave (GW) observations~\cite{Znajek}. 
These dissipation mechanisms have roots in the large charge-to-mass-ratio of the electron~\cite{2016JCAP...05..054C}.
As a consequence, gravitational-wave (GW) searches, parameter estimation (PE), and population studies~\cite{O3a_catalog, Abbott:2020gyp} are routinely carried out by \textit{assuming} that BHs giving rise to the signals observed in the LIGO-Virgo interferometers~\cite{AdvLIGO, AdvVirgo} can be accurately described by the Kerr metric. 
Even though all these arguments rely on well understood physical principles and thus in standard astrophysical scenarios the neutral BH approximation is a reasonable one, a robust and direct observational verification of charge neutrality for the population of BHs observed by LIGO and Virgo is still missing.

The confirmation of small or null electrical charge would also constrain more exotic scenarios, where the charge parameter of the KN family can be identified with the magnetic charge (due to primordial magnetic monopoles~\cite{monopoles, Bozzola:2020mjx, Liu:2020cds}), vector charge in theories mediated by a gravitational vector field~\cite{Bozzola:2020mjx}, a hidden electromagnetic charge in models of minicharged dark matter~\cite{2016JCAP...05..054C}, or a topologically induced charge~\cite{Kim:2020bhg}. 
Models of minicharged dark matter would evade the aforementioned discharge mechanisms due to their different charge-to-mass-ratio, while charge effects arising from modified gravity scenarios would be due to the presence of an additional gravitational field. 
Given that at the scale of BH mergers all these effects can be parametrised with the same parameter appearing in Einstein-Maxwell theory, we will simply refer to this parameter as \textit{charge}, bearing in mind its different meanings depending on the context in which this parameter is interpreted.
Charged BHs have also recently gained interest as a possible explanation of ultra high energy cosmic ray particles~\cite{Liberati:2021uom}.
GWs constitute a unique probe of these exotic scenarios for stellar-mass binary black hole (BBH) mergers, since the corresponding electromagnetic (EM) signal emitted by such sources would lie in the kHz range, where plasma absorption and reflection by the interstellar medium would prevent the detection of an EM counterpart~\cite{2016JCAP...05..054C, Cardoso:2020nst}.
For these reasons, we will not discuss possible EM counterparts to GWs that would be present if BHs possess a charge; in the recent past this topic received considerable attention due to a putative EM counterpart to GW150914~\cite{Liebling:2016orx, Loeb:2016fzn, Fraschetti:2016bpm, Liu:2016olx, Zhang:2016rli}.

Finally, in addition to probing effects due to new physics or uncommon astrophysical scenarios, KN BHs provide an excellent opportunity to test current phenomenological paradigms to search for violations of the Kerr hypothesis in a plausible and well understood scenario. 
The KN case is in fact an extensively studied modification to Kerr BHs in GR, stemming from a well-posed extension of Einstein's equations, the Einstein-Maxwell theory. It will also include some of the effects one would find in Einstein-Maxwell-dilaton theory, which is also well-posed; see~\cite{Hirschmann:2017psw} for simulations of binary black holes in the theory starting from approximate initial data, \cite{Khalil:2018aaj,Julie:2018lfp} for post-Newtonian calculations, and~\cite{Ferrari:2000ep} for computations of quasinormal mode (QNM) frequencies for nonspinning black holes in this theory.
Conversely, most alternative theories of gravity which could likely leave detectable imprints in GW signals from binary BHs, are instead often not known to have a well-posed formulation or their effects on observable quantities have been computed only approximately, e.g.,~\cite{Yunes:2009ke, Nair:2019iur, Perkins:2021mhb, Wagle:2021tam, Pierini:2021jxd, Okounkova:2019dfo,Okounkova:2019zjf,Okounkova:2020rqw, Shiralilou:2020gah, Shiralilou:2021mfl, Srivastava:2021imr}. However, see~\cite{Kovacs:2020pns,Kovacs:2020ywu} for well-posed formulations of some theories, though still assuming weak coupling, and~\cite{East:2020hgw, East:2021bqk} for initial numerical simulations using these formulations.

Investigating the impact of the KN scenario on GW measurements is of paramount importance to explore complications that may arise when considering non-perturbative beyond-GR effects in a self-consistent manner. 
These complications include the excitation of additional modes not present in GR, and the correlations among beyond-GR parameters and BH intrinsic parameters in the different phases of the coalescence.
The LIGO-Virgo Collaboration (LVC) routinely applies a battery of tests to GR~\cite{TGR-LVC2016, LIGOScientific:2019fpa, O3a_catalog} on confident GW detections, aimed at detecting deviations from GR predictions in the observed signals. 
A variety of effects are tested with different methodologies, including modifications to the generation or propagation of GWs, the nature of the merging objects, or the presence of additional polarizations, absent in GR. 
Residuals in the interferometer strain, obtained subtracting a representative best-fit waveform, are also tested for the presence of additional coherent power not modeled by GR templates~\cite{Ghonge:2020suv, O3a_TGR}.
Some of these tests are in principle sensitive to the presence of charge. Examples are the parametrised family of tests targeting the emission of dipole radiation during the early inspiral~\cite{Abbott:2018lct, LIGOScientific:2019fpa, O3a_TGR}, the parametrised ringdown tests~\cite{O3a_TGR, Brito:2018rfr, ghosh2021constraints, CARULLO-DELPOZZO-VEITCH, ISI-GIESLER-FARR-SCHEEL-TEUKOLSKY, Isi:2021iql} and the inspiral-merger-ringdown (IMR) consistency test~\cite{IMR_consistency_test1, IMR_consistency_test2}. 

Nevertheless, the aforementioned tests all follow a phenomenological approach, meaning that they do not assume a specific form for the modification to GR. 
The un-modelled approach is thus non-committal to a specific alternative scenario. On the one hand, this is a desirable feature, given the extraordinarily large number of possible alternatives to GR~\cite{Clifton:2011jh}. 
On the other hand, ignoring predictions from specific theories implies a loss in sensitivity when looking for deviations from the GR Kerr predictions.

In this work, we improve on the aforementioned agnostic tests and search for signatures of astrophysical or exotic charges in the merger-ringdown signal of BBH coalescences detected by the LIGO-Virgo interferometers. 
We do so by tabulating the QNM frequencies of a KN BH for arbitrary values of charge and spin, building on the work of Ref.~\cite{Dias:2015wqa}, and constructing a GW template implementing these predictions. 
We then use this template to perform an observational analysis on all confident post-merger BBH observations, deriving a bound on the maximum amount of charge compatible with current observations. Additionally, we present a study of the detectability of charge using the projected design sensitivity of the current detector network.
We employ a robust statistical framework and, for the first time, a non-perturbative treatment of the effects of charge and spin in the gravitational ringdown modes, without relying on assumptions such as a small-charge or WKB approximations, as used in previous analyses~\cite{2016JCAP...05..054C, Wang:2021uuh}.
We also take into account possible modifications in the amplitude of the waveform, in addition to the modifications to the phase. 

We additionally compute analytic fits for the QNM frequencies as a function of mass and spin. Such fits are a crucial ingredient for the construction of complete inspiral-plunge-merger-ringdown \textit{analytical} templates for charged binary black holes (generalizing the ones available for uncharged black holes) aimed at routinely extracting all the possible available information on BH charges from current LVC observations. Such templates will also require input from post-Newtonian calculations~\cite{Khalil:2018aaj,Julie:2018lfp} and numerical relativity simulations~\cite{Bozzola:2020mjx,Bozzola:2021elc} in the charged case.

The paper is structured as follows: in Sec.~\ref{sec:QNM_NR} we summarise the results obtained in a companion study~\cite{QNMsKN}, discussing a large dataset of QNM numerical solutions as a function of the charge and spin of the remnant BH.
In Sec.~\ref{sec:QNM_fits} we use the numerical data to construct parametric fits in an analytical form. We give the fit coefficients in the Appendix. Sec.~\ref{sec:LVC_DA} deals with the construction of a suitable waveform template describing a KN BH resulting from a BBH coalescence and the analysis of all available merger-ringdown observations from the LVC. 
Sec.~\ref{sec:Injections} discusses the prospects of extracting the BH charge from upcoming merger-ringdown observations using ground-based interferometers. Finally, we conclude and discuss future developments in Sec.~\ref{sec:Conclusions}.
Throughout the manuscript we use both $c=G=1$ units and Gaussian units in the electromagnetic sector. The charge to mass ratio, the parameter entering QNM computations directly, is defined by $\bar{q}:= |Q|/M$, with $Q$ the BH charge in Gaussian units, and the absolute value is quoted due to the QNM spectrum invariance with respect to the sign of the charge in the Einstein-Maxwell theory. Additionally, for statistical quantities, we quote the median and $90\%$ credible levels (CL) -- or credible regions (CR) when discussing multidimensional distributions -- unless explicitly stated otherwise. When discussing observations, we always quote BH redshifted masses, as observed in a geocentric reference frame.

\section{Numerical QNM computation}\label{sec:QNM_NR}

\begin{figure*}
\includegraphics[width=.46\textwidth]{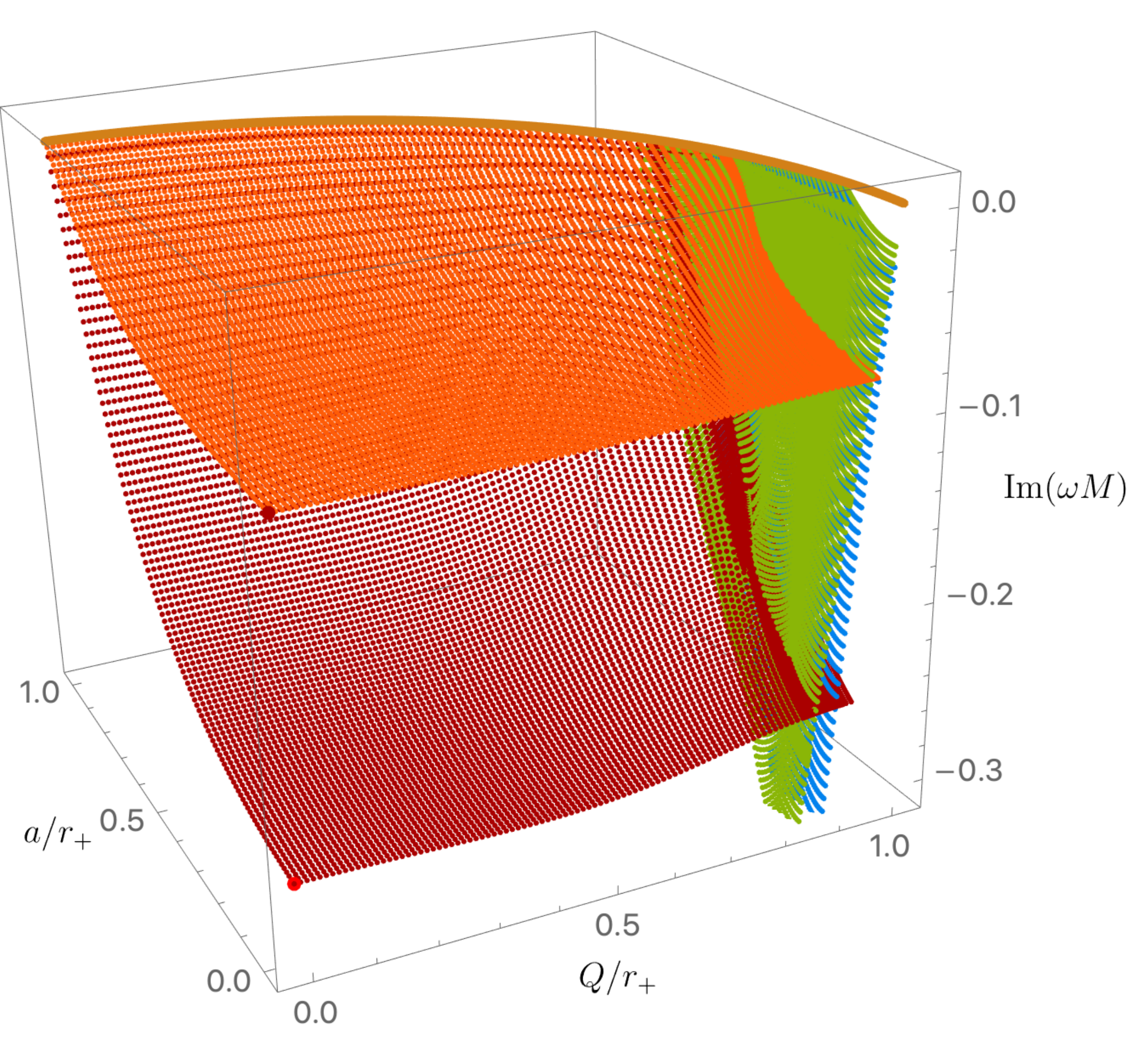}
\hspace{1.5cm}
\includegraphics[width=.44\textwidth]{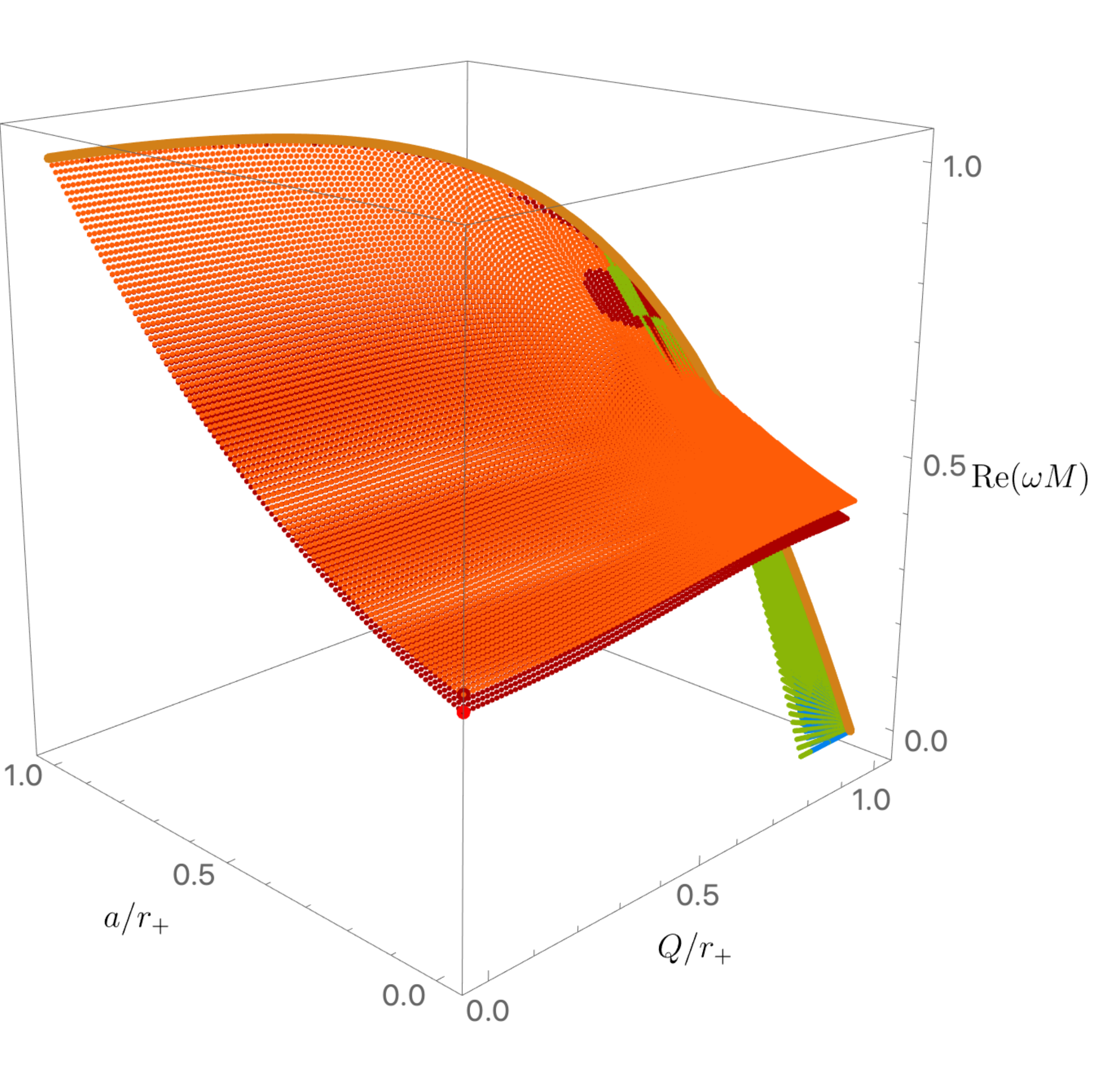}
\caption{Imaginary (left panel) and real (right panel) part of the frequency for the $Z_2$ $\ell=m=2$ KN QNMs. 
The orange surface (top of both figures) represents the $\mathrm{PS}$ family of the $n=0$ mode. The green surface (right side of both figures) represents the $\mathrm{NH}$ family of the $n=0$ mode.
The dark-red surface (below the orange surface) corresponds to the $\mathrm{PS}$ family of the $n=1$ mode.
Finally, the blue surface (almost on the top of the green surface) corresponds to the $\mathrm{NH}$ family of the $n=1$ mode.
The point at $a=0=Q$ in the orange (dark-red) surfaces matches the gravitational QNM of Schwarzschild [60,61] for the $n=0$ ($n=1$) modes, while the brown curve marks the extremal limit.
In these figures and the others we just plot the NH surface up to $Q/r_+=0.99$ which explains the small gap between the green (blue) surface and the extremal brown curve. These highly charged values can be computed with an analytical formula derived in~\cite{QNMsKN,ExtendedQNMsKN} that provides an excellent approximation to the numerical solution. We display $\mathrm{NH}_n$ (i.e., the green and blue surfaces) only for large charge where they can dominate over the $\mathrm{PS}_n$ sub-families; for smaller charge they are very strongly damped.
}\label{Fig:Z2l2m2n0n1-rp}
\end{figure*} 

In the 1980s, Chandrasekhar, in his seminal textbook~\cite{Chandra:1983}, identified the coupled system of two partial differential equations (PDEs) that, under a particular gauge choice, describe the most general gravito-electromagnetic perturbations of a KN BH (excluding the perturbations that change the mass, angular momentum and/or charge of the solution). See also Ref.~\cite{Mark:2014aja}.
Since then, different studies have attempted to solve these PDEs using certain approximations. In a first attempt,~\cite{Kokkotas_charge, Berti:2005eb} studied perturbations described by the so-called Dudley-Finley equations. This is a decoupled system of Teukolsky-like equations that describes exactly the spin $0$ (scalar field) perturbations of KN and are expected to be a reasonably good approximation for the higher spin gravito-electromagnetic perturbation equations. Later, Chandrasekhar's equations for KN were solved perturbatively in a small rotation ($a$) expansion about the Reissner-Nordstr\"om BH~\cite{Pani:2013ija,Pani:2013wsa}, and in a small charge ($Q$) expansion about the Kerr background~\cite{Mark:2014aja}. The calculation of QNMs, within a WKB and/or near-horizon approximation, in the KN extremal limit was also considered in Ref.~\cite{Zimmerman:2015trm}.

More recently, it was shown that the most general gravito-electromagnetic perturbations of KN can be described by a coupled system of two PDEs 
for two  gauge invariant Newman-Penrose fields~\cite{Dias:2015wqa} that, 
upon gauge fixing, reduce to the PDE system originally found by Chandrasekhar~\cite{Chandra:1983,Mark:2014aja}. These equations were then solved numerically using numerical methods reviewed in~\cite{Dias:2015nua}, in relevant ranges of the KN parameter space (notably for KN with $a=Q$), to establish strong evidence in favour of linear mode stability of the KN BH against gravito-electromagnetic perturbations \cite{Dias:2015wqa} (further supported by the non-linear time evolution study of~\cite{Zilhao:2014wqa}). 
However, it was only recently~\cite{QNMsKN,ExtendedQNMsKN}, that the most desired task of obtaining the full frequency spectra of the QNMs with slowest decay rate (and others of physical interest) was finally completed. In the rest of this section, we will borrow and discuss results from our companion papers~\cite{QNMsKN,ExtendedQNMsKN} that will form the theoretical basis for the present study.

\begin{figure*}
\includegraphics[width=.46\textwidth]{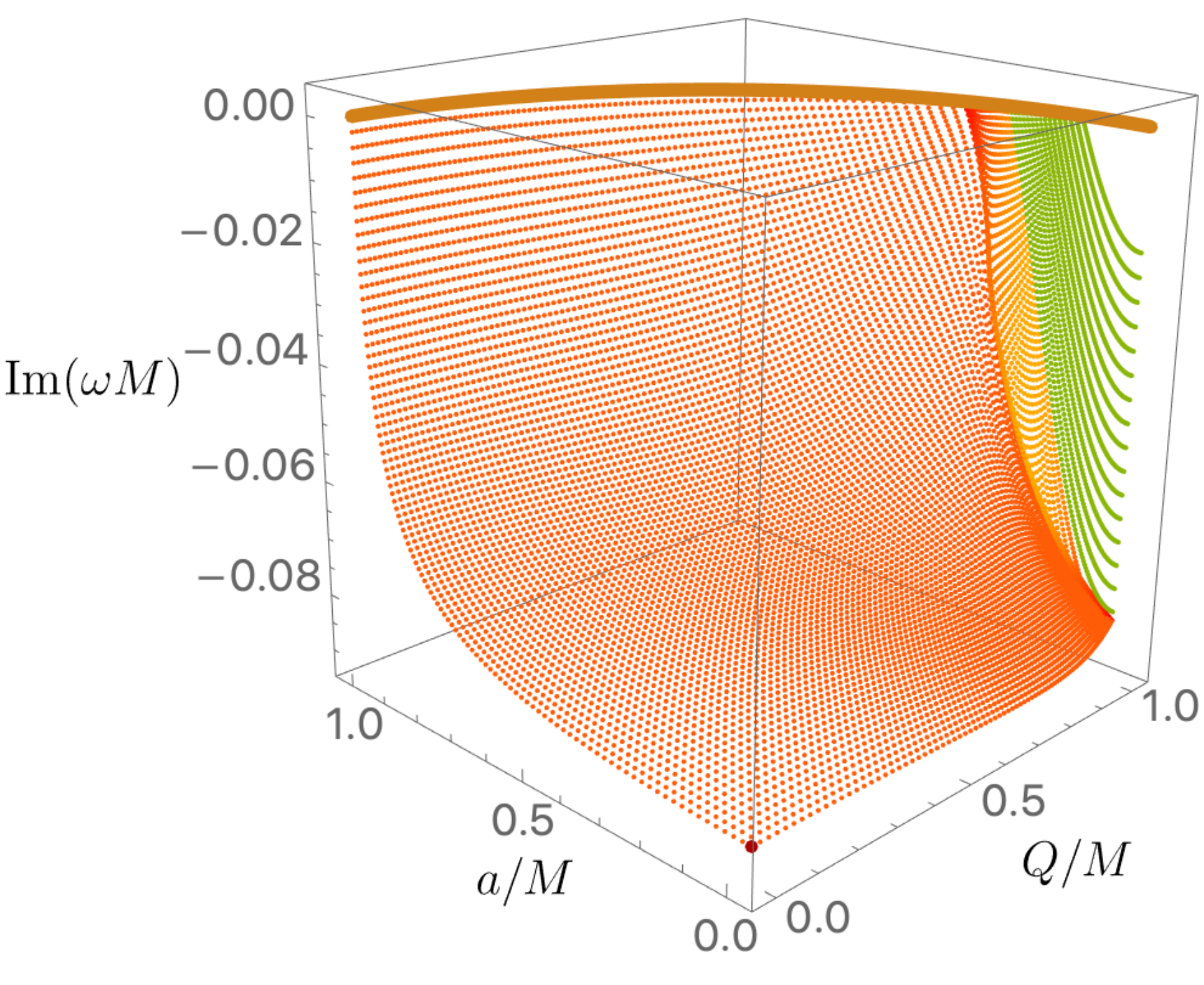}
\hspace{1.5cm}
\includegraphics[width=.44\textwidth]{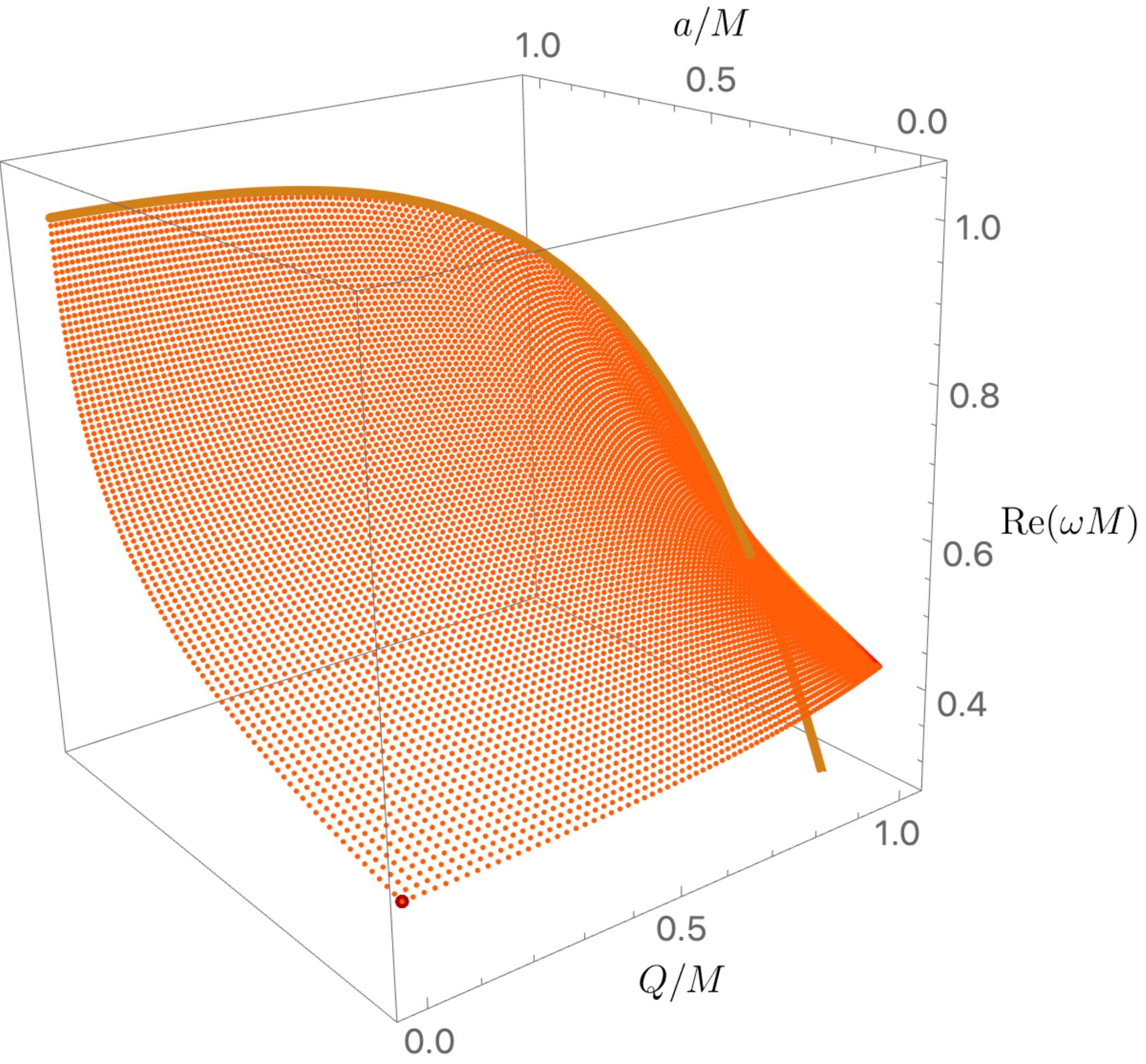}
\caption{Imaginary (left panel) and real (right panel) parts of the frequency for the $Z_2$, $\ell=m=2, n=0$ KN QNM with lowest $\mathrm{Im}\,|\hat{\omega}|$.
At each ($a/M$, $q/M$) point, only the ``dominant'' QNM family (i.e., the one with the larger damping time between the PS and NH families) is shown. 
The orange (green) surface denotes the region where the PS (NH) family is dominant.
The yellow area indicates the region where the two families of modes trade dominance.
At extremality, the dominant mode always starts at $\mathrm{Im}\,\hat{\omega}=0$ and $\mathrm{Re}\,\hat{\omega}=m\hat{\Omega}_H^{\hbox{\footnotesize ext}}$ (brown curve). 
The dark-red point ($a=0=Q$), $\hat{\omega}\simeq 0.37367168 - 0.08896232\, i $, is the gravitational QNM of Schwarzschild ~\cite{Chandra:1983,Leaver:1985ax}. 
In the right panel, the yellow and green regions are so close to the extremal brown curve that they are not visible.}
\label{Fig:Z2l2m2n0}
\end{figure*}  

\begin{figure*}
\includegraphics[width=.45\textwidth]{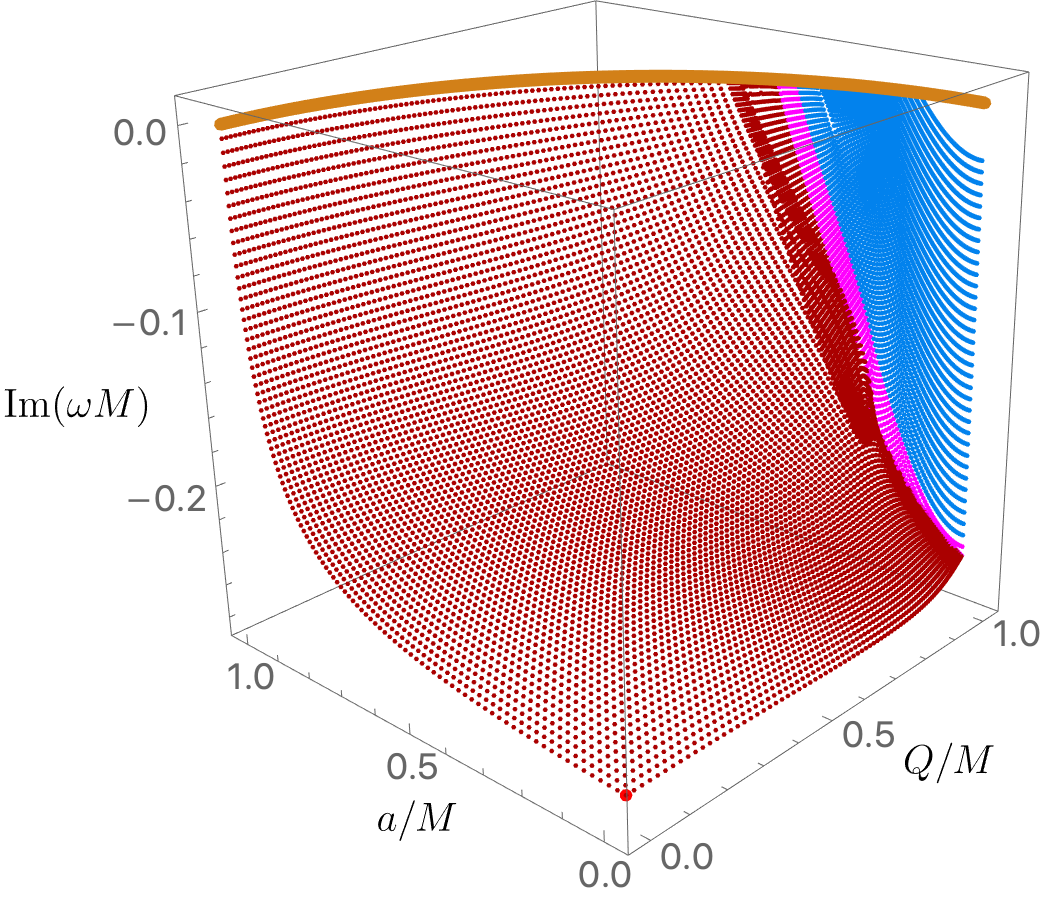}
\hspace{1.5cm}
\includegraphics[width=.45\textwidth]{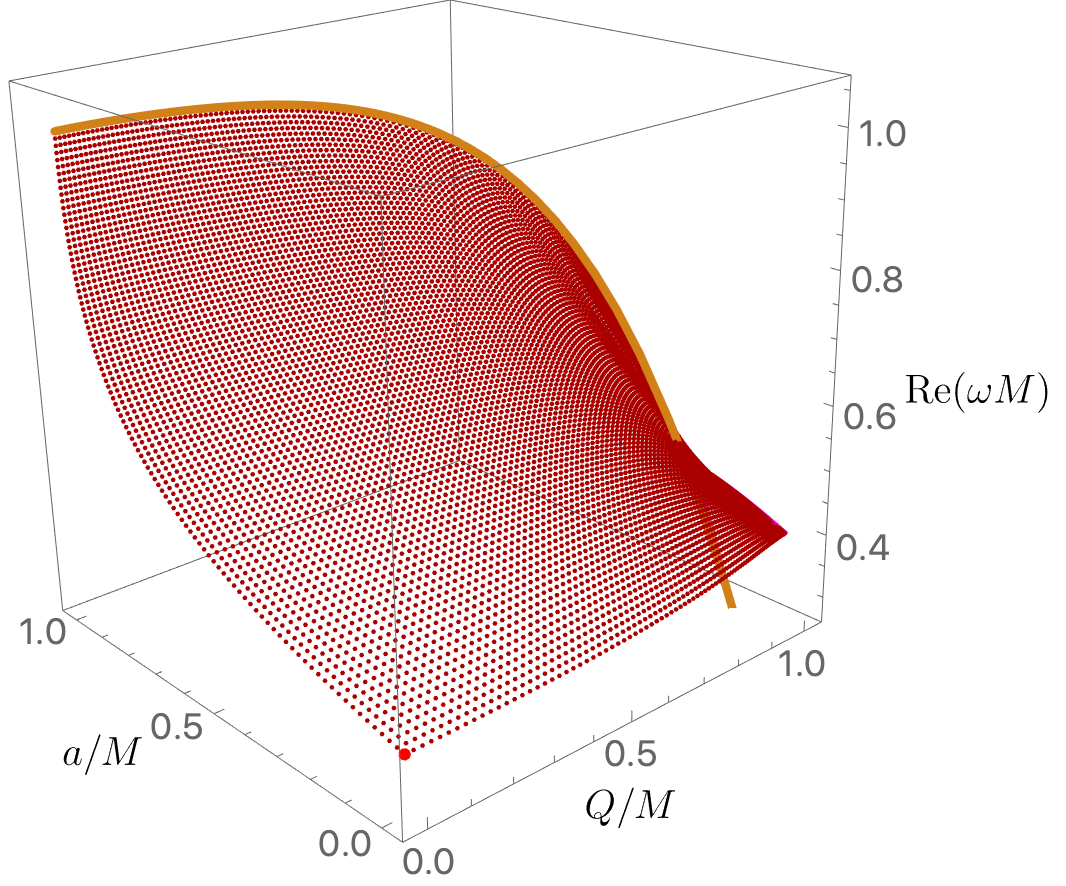}
\caption{Imaginary (left panel) and real (right panel) parts of the frequency for the $Z_2$, $\ell=m=2, n=1$ KN QNM with lowest $\mathrm{Im}\,|\hat{\omega}|$.
At each ($a/M$, $q/M$) point, only the ``dominant'' QNM family (i.e., the one with the larger damping time between the PS and NH families) is shown. 
The dark-red (blue) surface denotes the region where the PS (NH) family is dominant.
The magenta area indicates the region where the two families of modes trade dominance.
At extremality, the dominant mode always starts at $\mathrm{Im}\,\hat{\omega}=0$ and $\mathrm{Re}\,\hat{\omega}=m\hat{\Omega}_H^{\hbox{\footnotesize ext}}$ (brown curve).
The red point ($a=0=Q$), $\hat{\omega}\simeq 0.34671099 - 0.27391488\, i $, is the gravitational QNM of Schwarzschild ~\cite{Chandra:1983,Leaver:1985ax}. 
In the right panel, the magenta and blue regions are so close to the extremal brown curve that they are not visible.}
\label{Fig:Z2l2m2n1}
\end{figure*}  

\begin{figure*}
\includegraphics[width=.45\textwidth]{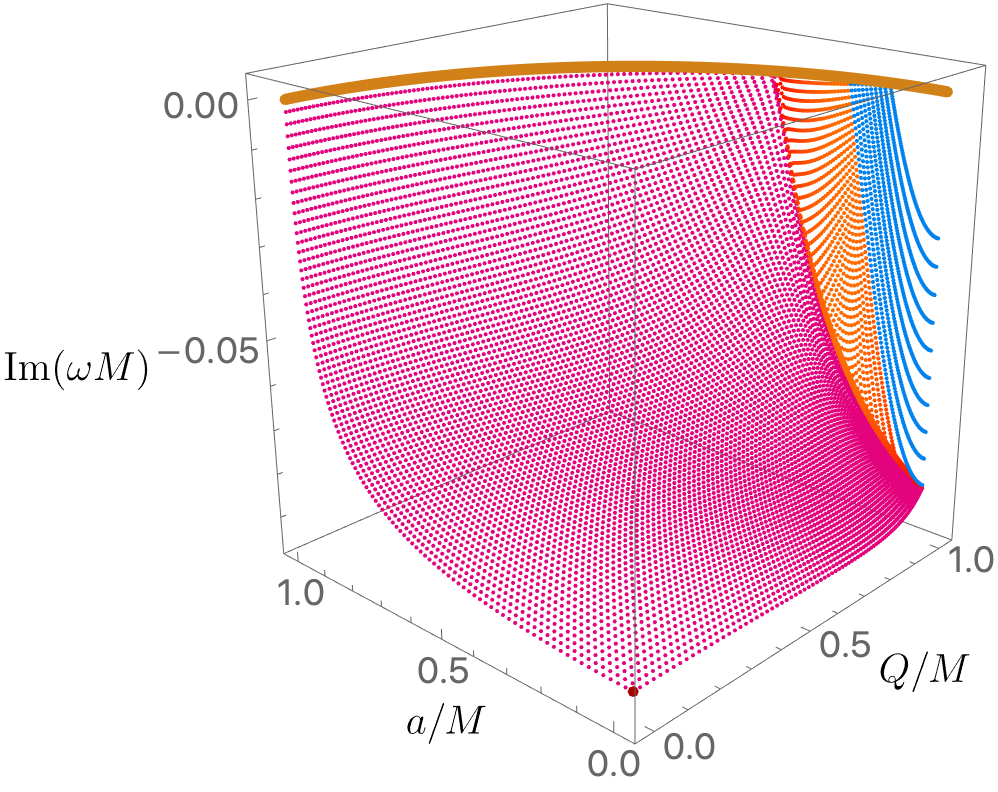}
\hspace{1.5cm}
\includegraphics[width=.45\textwidth]{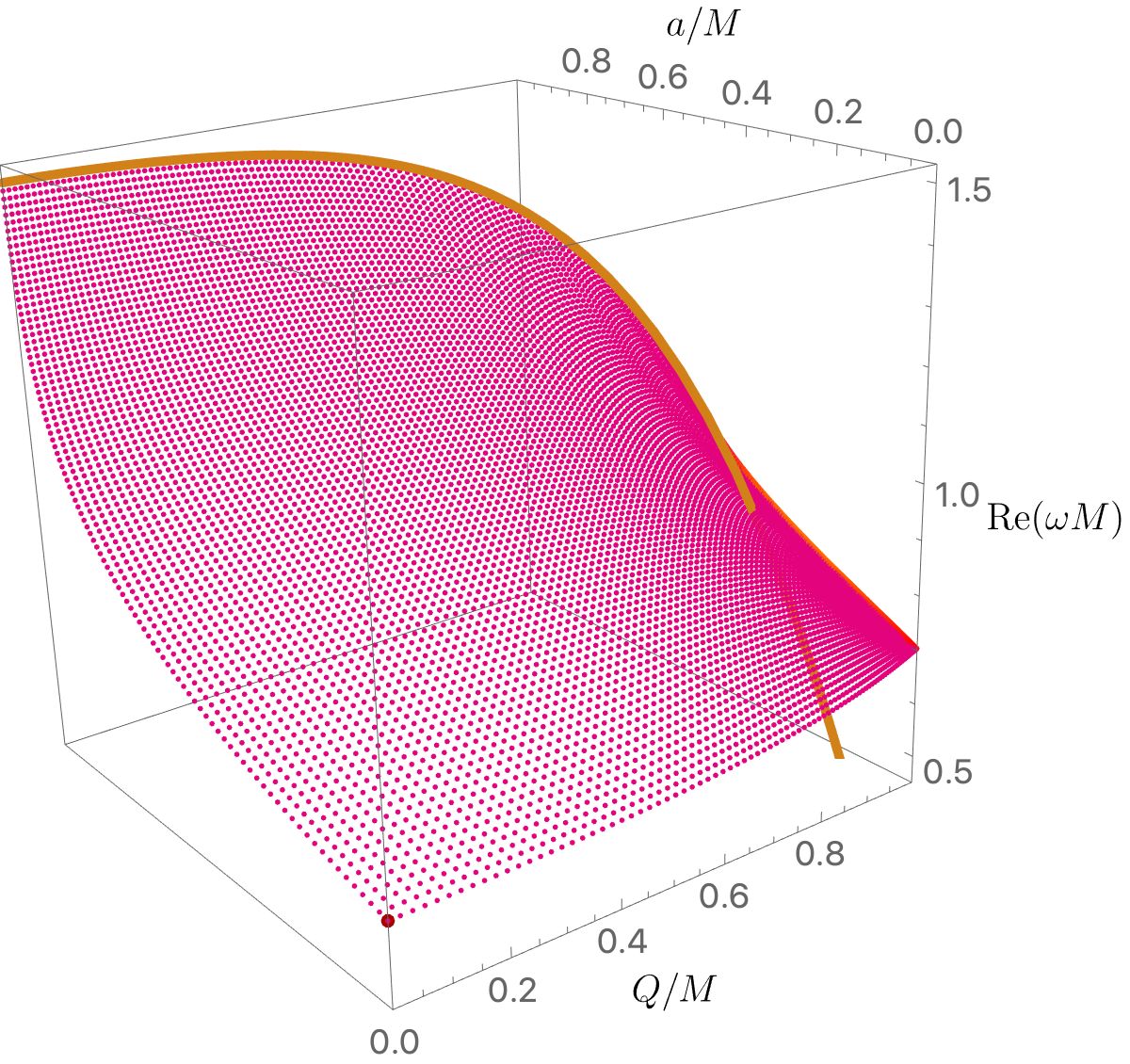}
\caption{Imaginary (left panel) and real (right panel) parts of the frequency for the $Z_2$, $\ell=m=3, n=0$ KN QNM with lowest $\mathrm{Im}\,|\hat{\omega}|$.
At each ($a/M$, $q/M$) point, only the ``dominant'' QNM family (i.e., the one with the larger damping time between the PS and NH families) is shown. 
The magenta (blue) surface denotes the region where the PS (NH) family is dominant.
The orange area indicates the region where the two families of modes trade dominance. 
At extremality, the dominant mode always starts at $\mathrm{Im}\,\hat{\omega}=0$ and $\mathrm{Re}\,\hat{\omega}=m\hat{\Omega}_H^{\hbox{\footnotesize ext}}$ (brown curve). 
The dark-red point ($a=0=Q$), $\hat{\omega}\simeq 0.59944329 - 0.09270305\, i $, is the gravitational QNM of Schwarzschild ~\cite{Chandra:1983,Leaver:1985ax}.
In the right panel, the orange and light-blue regions are so close to the extremal brown curve that they are not visible.}
\label{Fig:Z2l3m3n0}
\end{figure*}  

For the astrophysical investigations considered in this work, we wish to identify the families of gravito-electromagnetic QNMs that dominate the ringdown emission following a BBH merger, focusing on the perturbations with spin weight $-2$. 
Here by dominant we mean the families that have the slowest decay rates for all KN BHs parametrized by the $\{a,Q\}$ pairs\footnote{We use the notation of~\cite{QNMsKN,ExtendedQNMsKN}. In Boyer-Lindquist coordinates, the outer and inner horizon radii $r_\pm$ are related to the KN mass $M$ and charge $Q$ by $r_\pm=M\pm\sqrt{M^2-a^2-Q^2}$ and the event horizon angular velocity and temperature are  $\Omega_H= \frac{a}{r_+^2+a^2}$ and $T_H = \frac{1}{4 \pi  r_+}\frac{r_+^2-a^2-Q^2}{r_+^2+a^2 }$. At $r_-=r_+$, i.e., $a=a_{\hbox{\footnotesize ext}}=\sqrt{M^2-Q^2}$, the KN BH has a regular extremal (``ext") configuration with $T_H^{\hbox{\footnotesize ext}} =0$, and maximum angular velocity $\Omega_H^{\hbox{\footnotesize ext}} =a_{\hbox{\footnotesize ext}}/(M^2+a_{\hbox{\footnotesize ext}}^2)$.}.
Not surprisingly, the dominant family of QNMs is the one that, in the $a=Q=0$ limit and using Chandrasekhar's notation \cite{Chandra:1983}, reduces to the Schwarzschild gravitational $Z_2$ $\{ \ell=m=2, n=0 \}$ mode. Here $\ell$ is the harmonic number that gives the number of zeros of the QNM eigenfunction along the polar direction and $n$ is the radial overtone (related to the number of zeros of the QNM eigenfunction along the radial direction).  
The second family of interest is the one that reduces to the gravitational $Z_2$ $\{ \ell=m=2, n=1 \}$ mode in the Schwarzschild limit~\cite{Chandra:1983}. Although this mode has a short lifetime, in the uncharged case it contributes significantly to the emission soon after the peak of the GW waveform~\cite{Giesler:2019uxc} due to its large excitation. These QNM spectra were obtained in our companion paper \cite{QNMsKN} and further detailed in the associated extended study~\cite{ExtendedQNMsKN}.
Finally, we will also need information about the spectra of the QNM family that reduces to the Schwarzschild gravitational $Z_2$ $\{ \ell=m=3, n=0 \}$ mode in the $a=Q=0$  limit. This mode makes a significant contribution to the emission for BBH mergers where the progenitor's mass ratio is significantly different from unity.

The task of identifying the most dominant modes within each of these families of QNMs is made less trivial by the fact that for each family $\{\ell=m,n\}$ there are not one but actually two sub-families of QNMs \cite{QNMsKN,ExtendedQNMsKN}. These can be denoted as
1) the {\it photon sphere} ($\mathrm{PS}_n$), and 2) the {\it near-horizon} ($\mathrm{NH}_n$) sub-families, although this sharp distinction is unambiguous only for small rotation parameters, i.e., close to the Reissner-Nordstr\"om family. To classify them in the Kerr-Newman background, we start by identifying them in the Reissner-Nordstr\"om limit and then we follow these two sub-families as the rotation parameter increases.
In this Reissner-Nordstr\"om case, the PS family of QNMs is the one that in the eikonal or geometric optics limit $-$ i.e. the WKB limit  $m=\ell\to \infty$ $-$ has a frequency spectrum that is closely connected to the properties of unstable equatorial circular photon orbits: the real and imaginary parts of the PS frequency are proportional to the Keplerian frequency and to the Lyapunov exponent of the orbit, respectively. The latter describes how quickly radial deformations increase the cross section of a null geodesic congruence around the orbit. On the other hand, the NH family of QNMs is characterized by having a wavefunction that near-extremality is very much localized around the horizon and quickly decays to zero as we move away from it. It is further characterized by the fact that its QNM spectrum has an imaginary part that vanishes in the extremal limit and, in the Reissner-Nordstr\"om case, has vanishing real part (unlike the PS modes).
Starting in the Reissner-Nordstr\"om solution, as the rotation increases and we run over the KN parameter space, these PS and NH sub-families define two surfaces (in a $\{Q,a,\omega\}$ plot) that do intersect (with a simple crossover) or have the interesting phenomena of eigenvalue repulsions in the KN parameter space as detailed in~\cite{QNMsKN,ExtendedQNMsKN}. Typically, this happens for very large values of $Q/M$, in a region of the parameter space which is difficult to probe with observations. When eigenvalue repulsion occurs, instead of a simple intersection, it becomes harder to make a clear distinction between the PS and NH sub-families. 
Additional difficulties emerge from the fact that non-trivial intersections with eigenvalue repulsions can also happen between different (sub-)families, e.g., between $\{\ell=m=2,n=0\}$ and $\{\ell=m=2,n=1\}$ modes.
This requires a careful analysis of the data to identify to which (sub-)family a particular QNM belongs; see Ref.~\cite{QNMsKN,ExtendedQNMsKN} for additional details.
For each KN BH parametrized by $\{a,Q\}$, we derive all the relevant $\mathrm{PS}_n$ and $\mathrm{NH}_n$ QNMs for a given $\{\ell=m,n\}$ family, and identify the modes that have the slowest decay rate within that particular $\{\ell=m,n\}$ QNM family.

An overview of four (out of the six) sub-families of QNMs that we need for our study is presented in Fig.~\ref{Fig:Z2l2m2n0n1-rp}, in order to give a general reference of their relative position in the frequency plane. Namely, in this figure we focus on the $\{\ell=m=2,n=0\}$ and $\{\ell=m=2,n=1\}$ families and, for each of them, we display the spectra of each of their two  sub-families, namely $\mathrm{PS}_n$ and $\mathrm{NH}_n$.
We plot the imaginary (left panel) and real (right panel) part of the frequency for these KN QNMs as a function of the KN rotation and charge.
In this particular figure we use dimensionless quantities in units of the horizon radius $r_+$ (instead of units of $M$), namely $\hat{a}:= a/r_+, \hat{Q}:= Q/r_+$, because it turns out that the distinction between the four sub-families is better seen in these units. For the frequency we always use units of $M$: $\hat{\omega} :=\omega M$, $\hat{\Omega} := \Omega M$. The brown curve with $\mathrm{Im}\,\hat{\omega}=0$ and  $\mathrm{Re}\,\hat{\omega}=2\hat{\Omega}_H^{\mathrm{ext}}$ corresponds to extremality (`ext') where $\hat{a}_\mathrm{ext}=\sqrt{1-\hat{Q}^2}$.
Note that, as expected, $\mathrm{PS}_{0}$ always has smaller $|\mathrm{Im}\,\hat{\omega}|$ than $\mathrm{PS}_{1}$, and $\mathrm{NH}_{0}$ always has smaller $|\mathrm{Im}\,\hat{\omega}|$ than $\mathrm{NH}_{1}$. 
Focusing our attention on the families with slowest decay rate, the $\mathrm{PS}_{0}$ and $\mathrm{NH}_{0}$ curves intersect at large charge (with simple crossovers or with intricate eigenvalue repulsions not clear in this overview figure, but identified in \cite{QNMsKN,ExtendedQNMsKN}).
Starting from $Q=0$ until a critical large charge $\hat{Q}=\hat{Q}_c(\hat{a})$, the $\mathrm{PS}_{0}$ dominates the QNM spectra and terminates on the brown curve at extremality, while for $\hat{Q}_c(\hat{a})<\hat{Q}\leq \hat{Q}_{\mathrm{ext}}$ it is instead the $\mathrm{NH}_{0}$ (which always terminates at the brown curve) that has the slowest rate. It should be noted that the $\mathrm{NH}_{0}$ green surface also intersects the $\mathrm{PS}_{1}$ dark-red surface often with eigenvalue repulsions that are not clearly visible in Fig.~\ref{Fig:Z2l2m2n0n1-rp} but that are detailed in \cite{QNMsKN,ExtendedQNMsKN}. They also leave an imprint in the dark-red $\mathrm{PS}_1$ surface that is partially visible in the left panel of Fig.~\ref{Fig:Z2l2m2n1} just to the left of the magenta region.
For reference, although not shown in Fig.~\ref{Fig:Z2l2m2n0n1-rp}, the $\mathrm{PS}_0$ surface of the $\{\ell=m=3,n=0\}$ QNMs would be in between the orange and dark-red surfaces, and the $\mathrm{NH}_0$ surface of the $\{\ell=m=3,n=0\}$ QNMs would be in between the green and blue surfaces.

After this generic overview of two of the main QNM families of interest for our study, we now give the QNM spectra of the {\it slowest decaying mode} for each of the three main families, namely $\{\ell=m=2,n=0\}$, $\{\ell=m=2,n=1\}$ and $\{\ell=m=3,n=0\}$, used in our study. This time we parametrize the KN BH by $\chi= a/M, \bar{q}= Q/M$.

In Fig.~\ref{Fig:Z2l2m2n0}, we show the result for the $\{\ell=m=2,n=0\}$ mode. For small and intermediate charge, the spectra is dominated by the $\mathrm{PS}_{0}$ mode (orange surface). On the other hand for very large charge (up to $Q/M=1$), the spectra is instead dominated by the $\mathrm{NH}_{0}$ mode (green surface). In between these two, there is a yellow area in the left panel where the $\mathrm{PS}_{0}$ and $\mathrm{NH}_{0}$ intersect either with a simple crossover or eigenvalue repulsions and they trade dominance. The yellow area picks the frequency of the mode that has the smallest $|\mathrm{Im}\,\hat{\omega}|$. In the right plot of Fig.~\ref{Fig:Z2l2m2n0}, the yellow and green areas are not visible because in units of $M$ the real part of their frequency is very, very close to the extremal brown curve.\footnote{A similar discussion applies to Figs.~\ref{Fig:Z2l2m2n1} and \ref{Fig:Z2l3m3n0}.}
These surfaces are however well visible when we use horizon radius units: see  
Fig.~\ref{Fig:Z2l2m2n0n1-rp}.

In Fig.~\ref{Fig:Z2l2m2n1}, we repeat the exercise for the $\{\ell=m=2,n=1\}$ mode. The $\mathrm{PS}_{1}$ mode (dark-red) dominates for small and intermediate charges, while the $\mathrm{NH}_{1}$ mode (blue) dominates for very large charges. In between, there is a small window with a magenta area where the $\mathrm{PS}_{1}$ and $\mathrm{NH}_{1}$ modes trade dominance and we display the mode with smallest $|\mathrm{Im}\,\hat{\omega}|$. In the rightmost side of the dark-red $\mathrm{PS}_{1}$ surface one identifies a small region where the surface is very deformed by the eigenvalue repulsion between this $\ell=m=2$ $\mathrm{PS}_{1}$ family and the $\ell=m=2$ $\mathrm{NH}_{0}$ family as detailed in \cite{QNMsKN,ExtendedQNMsKN}. 

Finally,  in Fig.~\ref{Fig:Z2l3m3n0} we give the spectra for the $\{\ell=m=3,n=0\}$ mode. The  $\mathrm{PS}_{0}$ mode (magenta surface) dominates for small and intermediate charges.
The light-blue surface is the $\mathrm{NH}_{0}$ mode and dominates for very large charges. In between, the orange area describes the region where the $\mathrm{PS}_{0}$ and $\mathrm{NH}_{0}$ modes trade dominance and we show the mode with smallest $|\mathrm{Im}\,\hat{\omega}|$.

\section{Fitting formulae for the numerical solutions}\label{sec:QNM_fits}

In this section, after introducing our fitting algorithm and testing it on previous Kerr results, we construct analytical fits for the real and imaginary part of KN PS QNM frequencies as a function of the BH charge and spin parameters.

\subsection{Bayesian fitting method} 
We formulate the problem in the language of Bayesian inference, an extension of classical logic in the absence of complete information~\cite{Jaynes2003}.
Our fitting templates will be characterised by a set of coefficients, collectively labeled by $\vec{\theta}$, possibly different for each analytical form chosen. We infer the optimal (best-fit) values and related uncertainties for the coefficients by computing their probability distribution, the \textit{posterior distribution} $p(\vec{\theta} | d, \mathcal{H}, I)$, conditioned on the available numerical data $d$. The distribution is obtained through Bayes' theorem:

\be
    p(\vec{\theta} | d, \mathcal{H}, I) = \frac{p(\vec{\theta} | \mathcal{H}, I) \cdot p(d | \vec{\theta}, \mathcal{H}, I)}{p(d | \mathcal{H}, I)} \,, 
\ee
where $\mathcal{H}$ constitutes the parametric model describing the data (hypothesis), while $I$ denotes all the available background information.
The distribution $p(\vec{\theta} | \mathcal{H}, I)$ is the \textit{prior distribution}, encoding all the available information on the coefficients before the start of the inference process (e.g., the bounds within which we allow the coefficients to vary). 
If no \emph{a priori} information is available, the prior can be chosen to be uniform on $\vec{\theta}$ within a given range of interest. 
The last key ingredient in the numerator is the \textit{likelihood} function $p(d | \vec{\theta}, \mathcal{H}, I)$, which is fixed by the error distribution of the numerical data.
For the numerical fits discussed in this section, we assume a likelihood given by a zero-mean gaussian distribution with a standard deviation equal to the numerical uncertainty, together with uniform priors on the template coefficients.
The overall normalisation $\mathcal{Z} := p(d | \mathcal{H}, I)$, known as the \textit{evidence}, encodes the probability that the data $d$ can be described by the chosen model. 
This approach allows one to compute the full multi-dimensional probability distribution of the coefficient set, improving upon uni-dimensional error estimates on each of the coefficients, and avoiding convergence issues in highly dimensional problems.
To explore the posterior probability distribution we use the nested sampling~\cite{skilling2006} algorithm \texttt{CPNest}~\cite{cpnest}. 

\subsection{The Kerr case} 
\begin{figure*}[!tb]
\includegraphics[width=0.85\textwidth]{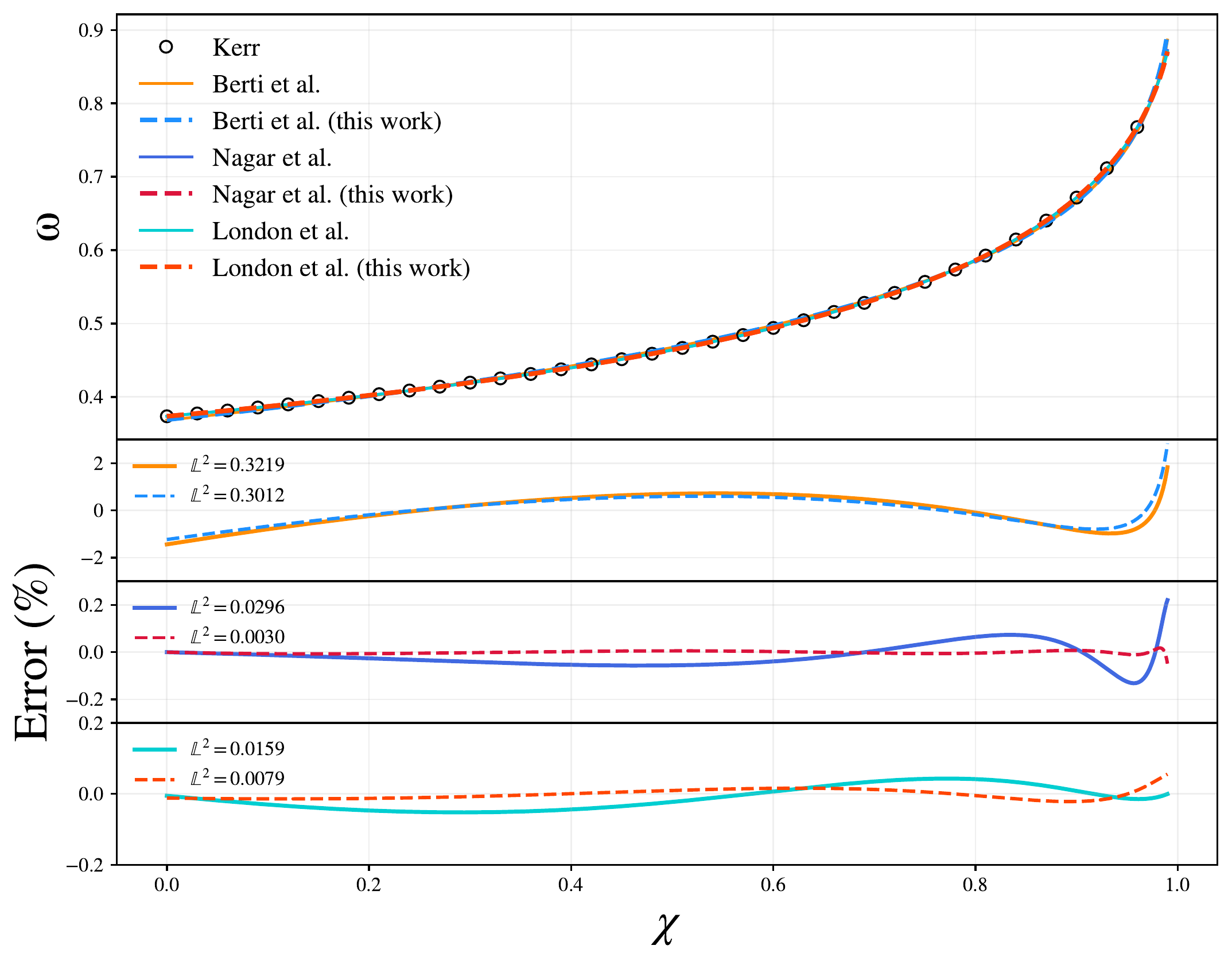}
\caption{Test of the fitting method (with the maximum posterior point estimate for the fit parameters) against results from the literature for the Kerr $(\ell,m,n)=(2,2,0)$ frequency, using the same ansatzs as in the original works. The fractional error is computed as the residual $(\omega^\text{data} - \omega^\text{fit}) / \omega^\text{data}$. The $\mathbb{L}^2$ norm quantifies the overall agreement with the data set. The open circles representing the numerical data have been down-sampled for visualisation purposes.}
\label{fig:Kerr_fits}
\end{figure*}

Before constructing a QNM template to model the KN case, we first test our fitting procedure by reproducing known results from the literature. As test cases, we choose the models of Berti et al.~\cite{Berti_fits}, Nagar et al.~\cite{Nagar_fits}, and London et al.~\cite{London_fits}. We start from the widely used analytical representation of Kerr BH spectra as a function of the BH spin from Berti et al.~\cite{Berti_fits}. It has the general form:
\be\label{eq:Berti_etal}
    X = c_0 +c_1 \cdot (1-\chi)^{c_2} \,, 
\ee
where $c_i \in \mathbb{R}$ and, defining the complex QNM frequency $\tilde{\omega} = \omega + i/\tau$, $X$ corresponds to $\omega$ or to the quality factor of each QNM mode, $Q= \omega \tau /2$.
The second model, employed in the construction of the effective one body model from Nagar et al.~\cite{Nagar_fits}, provides an improved representation of the spectrum with respect to Eq.~\eqref{eq:Berti_etal}, by assuming a rational function:
\be\label{eq:Nagar_etal}
    X = Y_0 \, \left( \frac{1+\sum_{j=1}^{3} b_j \, \chi^j}{1+\sum_{j=1}^{3} c_j \, \chi^j} \right) \,,
\ee
where $b_k, c_k \in \mathbb{R}$, $X$ corresponds to $\omega$ or $\tau^{-1}$ and $Y_0$ is the Schwarzschild value of the parameter under consideration.
The last model considered, from London et al.~\cite{London_fits}, has a precision comparable to the one of Eq.~\eqref{eq:Nagar_etal}, with the additional advantage of providing a smooth modeling of the near-extremal behaviour. 
It models directly the \textit{complex} QNM frequency by first smoothing the spectrum behaviour through a $\kappa$-transformation defined by:
\be
	\kappa := \left(\log_3(2-\chi)\right)^{\frac{1}{2+l-|m|}} \,, 
\ee
and subsequently modeling the QNM frequencies as:
\be\label{eq:London_etal}	
    \omega + i/{\tau} = \sum_{j=0}^{5} \kappa^j \, A_j \, e^{ip_j} \,, 
\ee
for each $(\ell,m,n)$ with $A_j\in \mathbb{R}, p_j\in [0, 2\pi]$.
We apply our fitting algorithm to the $(\ell,m,n)=(2,2,0)$ $\omega$ numerical data, publicly available from Ref.~\cite{Berti_website}, assuming each of the above templates, seeking to reproduce the results obtained in the original studies~\cite{Berti_fits,Nagar_fits,London_fits}.
Fig.~\ref{fig:Kerr_fits} shows the comparison of each template to the data both using the coefficients given in the original work and the ones obtained with our algorithm using the maximum \emph{a posteriori} values.
The fractional error is computed as the residual $(\omega^\text{data} - \omega^\text{fit}) / \omega^\text{data}$.
As expected, Eqs.~\eqref{eq:Nagar_etal} and \eqref{eq:London_etal} provide a more accurate description of the QNM frequency, with residuals around the $0.1\%$ level. Eq.~\eqref{eq:London_etal} proves to be the most faithful to the numerical data, especially in the extremal limit ($a \rightarrow$ 1). 
The overall agreement of each result with the data set employed is quantified by the $\mathbb{L}^2$ distance between the numerical data and the analytical formula. In the Nagar et al.\ case, our fits perform an order of magnitude better in terms of residuals and norm. However, it has to be noted\footnote{Alessandro Nagar, private communication.} that the fit of Nagar et al.\ was calibrated on values of the remnant spin corresponding to SXS catalog simulations employed in Ref.~\cite{Nagar_fits}. These include also some negative spin values, and only a subset of the data points for positive spin values considered here.
This dataset discrepancy might explain some of the difference in the residuals we observe. In the London et al.\ case, there is a general improvement in the non-extremal region, while the original fit provides a smoother behaviour around the extremality limit, although our result still shows a faithful representation below the $0.1 \%$ residual level for all the considered values of spin. This can be explained by the fact that we choose to not fix the extremal limit to minimise the global residuals, contrary to what was done in the original fit. Since our aim is simply to validate our numerical algorithm, which is returning compatible results with the ones of the original works, we do not try to resolve the small differences we observe in the residuals of these last two models, which are probably due to the aforementioned discrepancies in training data sets or fitting choices.

\subsection{The KN case} 

We now turn to the task of building a parametric function capable of modeling the spectrum of the gravitational KN QNMs discussed in Sec.~\ref{sec:QNM_NR}, the ones reducing to the Schwarzschild QNMs in the non-spinning, uncharged case.
We initially considered a generalization of the model used in Refs.~\cite{Pani:2013ija, Pani:2013wsa} to fit the small-charge case, including higher-order terms. However, we found that this is ineffective at modeling the spectrum in the large charge limit.
We instead choose to fit the numerical data presented in Sec.~\ref{sec:QNM_NR}, using a generalisation of the Nagar et al.\ model, Eq.~\eqref{eq:Nagar_etal}:
\be\label{eq:Nagar_etal_ext}
	X = Y_0 \, \left(\frac{\sum_{k,j=0}^{N_\text{max}} b_{k,j} \, \chi^k \, \bar{q}^j}{\sum_{k,j=0}^{N_\text{max}} c_{k,j} \, \chi^k \, \bar{q}^j}\right) \,.
\ee
Here $X$ corresponds to $\omega$ or $\tau^{-1}$, $Y_0$ stands for the Schwarzschild value of the corresponding fitted quantity, and $b_{k,j}, c_{k,j} \in \mathbb{R}$, with $b_{0,0}=c_{0,0}=1$.
This template contains $2 \cdot (N_\text{max}+1)^2 - 2$ free coefficients, implying that already truncating the expression to the same order used in the Kerr case, $N_\text{max}=3$, the number of coefficients increases to $30$, versus the $6$ coefficients used in the uncharged case.
We apply the Nested Sampling algorithm as described above, choosing uniform priors $\mathcal{U}(-10,10)$ for all the coefficients appearing in Eq.~\eqref{eq:Nagar_etal_ext}, and setting $N_\text{max}=3$ to limit the number of free parameters. 
We restrict our attention to the QNMs of interest for analysing observational data, hence we only consider data points respecting the sub-extremality condition $\chi^2 + \bar{q}^2 < $ 0.99.
In this region, the PS family is always dominant (longer damping time) compared to the NH one. The NH family has damping times comparable (or larger) to the PS one only very close to the extremal regime. 
Thus, in what follows, we only consider PS QNMs. 

\begin{figure*}[!tb]
\includegraphics[width=0.95\textwidth]{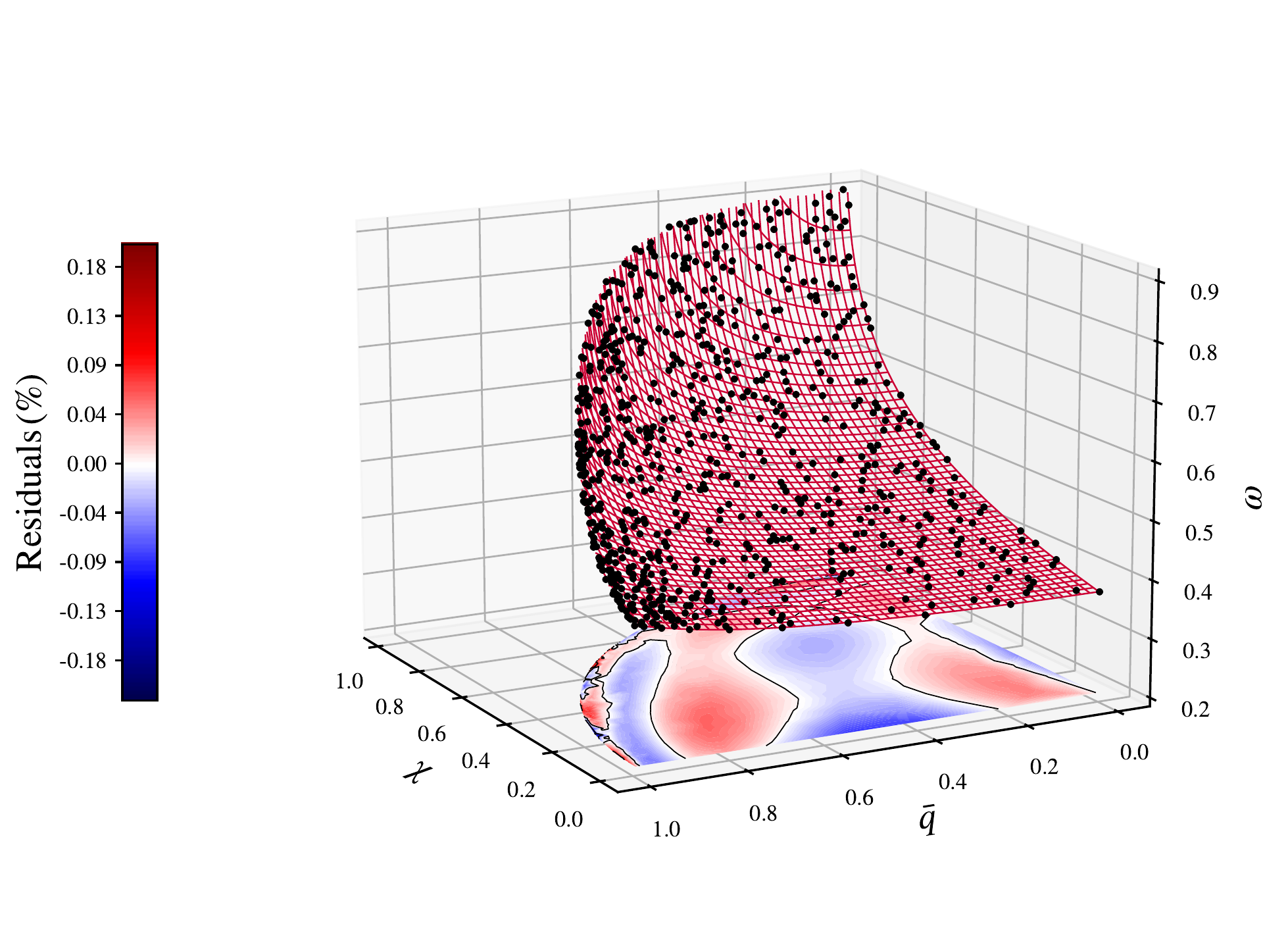}
\caption{Maximum \emph{a posteriori} fitting model for the $(\ell, m, n) = (2,2,0)$ QNM frequency, in red. Black points correspond to validation data points, excluded from the original fit, computed according to the methods described in Sec.~\ref{sec:QNM_NR}. The shadows encodes the corresponding residuals. The training dataset shows qualitatively identical behaviour, albeit with smaller residuals, as expected.}
\label{fig:omega_220_KN_fit}
\end{figure*}

We split the data into a training set, which constitutes $90\%$ of the full data set, and a validation set containing the remainder of the data. During the fit, we only employ the training set to find the values of the templates coefficients and use the validation set in a post-processing phase, to evaluate the residuals on values which were not used to construct the fit.
Fig.~\ref{fig:omega_220_KN_fit} shows the maximum a posteriori (which coincides with the maximum likelihood, since all priors are uniform) fitting model against the validation data points for the fundamental $(\ell, m, n) = (2,2,0)$ QNM frequency.
The residuals are centered around zero, spanning the range $\pm 0.2 \%$, indicating the same level of agreement of the best Kerr templates available.
We achieve comparable residuals on the frequencies and damping times of the other modes, except for the damping time of the  $(\ell, m, n) = (2,2,1)$ mode which shows residuals as high as $1\%$ in the corners of the parameter space, though the residuals drop below $0.5\%$ for $\chi^2 + \bar{q}^2 < 0.9$. This level of agreement is acceptable given the current and expected measurement precision obtainable on the damping time~\cite{O3a_TGR}.
The maximum of the posterior for both the frequency and damping time coefficients of the fits for the $(\ell, m,n) = \{ (2,2,0), (2,2,1), (3,3,0) \}$ modes, together with the median and $90\%$ CL on these coefficients are reported in the Appendix. As expected, residuals on the training set are of the same order of magnitude, although presenting smaller tails.

\section{Analysis of GW data}
\label{sec:LVC_DA}

In this section, after reviewing previous constraints and sensitivity predictions on BH charges from GW observations, we introduce our time-domain formalism and GW emission model used to infer the remnant object properties from GW data. This model is then applied to high confidence detections of GW transients with a sufficiently loud ringdown, presenting observational constraints on the maximum charge-to-mass ratio compatible with gravitational-wave data.

\subsection{Previous constraints on BH $U(1)$ charges}

Effects induced by the presence of a BH charge on GW signals were previously investigated in Refs.~\cite{2016JCAP...05..054C, Bozzola:2020mjx,Bozzola:2021elc, Wang:2020fra, Christiansen:2020pnv, Yunes:2016jcc, Wang:2021uuh, Gupta:2021rod} under different approximations. 
Ref.~\cite{Wang:2020fra} considered modifications to the early inspiral phase, although neglecting the effects of spins, and found no evidence for the presence of a BH charge. The impact of charge in the inspiral phase and the related PE were also analysed in Ref.~\cite{Christiansen:2020pnv} in simplified settings. 
Ref.~\cite{2016JCAP...05..054C} used the observation of GW150914 to place constraints on the dipolar emission, similarly to what was done in Ref.~\cite{Yunes:2016jcc} (for future prospects of constraining the dipolar emission using GW observations, see Refs.~\cite{Chamberlain:2017fjl, Perkins:2020tra}). A Bayesian study of the measurability of BH charges in the inspiral phase, considering the effects of charge up to first post-Newtonian order in the waveform phase was presented in Ref.~\cite{Gupta:2021rod}. This model was applied to GWTC-2 low-mass detections, providing the bound $\bar{q} < 0.2-0.3$ at $1$-$\sigma$ credibility.
A recent study on how some of these constraints from the inspiral phase could be affected by the presence of plasma surrounding the binary was presented in Ref.~\cite{Cardoso:2020nst}.
The detectability of charge in the ringdown emission was studied in Ref.~\cite{2016JCAP...05..054C} in the small charge limit, while recently Ref.~\cite{Wang:2021uuh} analysed the ringdown signal of GW150914 by including the effect of a BH charge using a WKB approximation. The results we find are in contrast with the bound obtained in this latter work. We attribute this difference to the KN spectrum of this latter reference being approximated using an ansatz based on the eikonal limit. This ansatz was further calibrated only on $\bar{q}=\chi$ numerical data, neglecting the full structure of the two-dimensional parameter space.
The limitations of the eikonal approximation for low-$\ell$ values (contributing to our analyses) are discussed in our companion paper~\cite{QNMsKN}.

Finally, a major step towards the full characterisation of waveforms sourced by KN metrics was taken in Refs.~\cite{Bozzola:2019aaw, Bozzola:2020mjx, Bozzola:2021elc}, where a set of complete numerical solutions of the inspiral, merger and ringdown of two charged non-spinning BHs in quasi-circular orbits was computed. The accuracy of different analytical approximations was evaluated against numerical results, pointing to a poor agreement of quantities estimated from a quadrupolar approximation in Newtonian models, while a much better agreement was found on remnant quantities estimates from the test particle limit.
The simulations were used to perform a mismatch analysis between charged and uncharged numerical solutions, allowing them to predict a constraint on the charge-to-mass-ratio of GW150914: $\bar{q} \leq 0.3$. 
This is the first prediction on the BH charge to stem from a full IMR comparison, although it has not been yet directly validated against observational data. The prediction was also obtained for a fixed mass ratio and neglecting the effect of spins, thus not taking into account the full correlation structure of the BBHs parameter space, an important point in an observational analysis, as will be discussed in the remainder of the paper.
The detectability predictions of Refs.~\cite{Bozzola:2020mjx, Bozzola:2021elc}, where applicable, are in good agreement with the results we obtain.

\subsection{Methods} 
\textit{pyRing --} We investigate the KN hypothesis in LIGO-Virgo data by employing the \texttt{pyRing}~\cite{CARULLO-DELPOZZO-VEITCH, ISI-GIESLER-FARR-SCHEEL-TEUKOLSKY, Carullo:2021dui} software, a \texttt{python}~\cite{python} package specifically tailored to the estimation of ringdown parameters. \texttt{pyRing} implements a Bayesian approach (see Sec.~\ref{sec:QNM_fits}), formulating the problem completely in the time domain, both for the likelihood and the waveform, in order to exclude any contribution from the pre-merger emission. Similarly to the numerical fits, the underlying stochastic sampling is performed by the \texttt{CPNest}~\cite{cpnest} algorithm. A convenient feature supported by the software is the possibility to generate synthetic data streams obtained by adding -- \emph{injecting}, in LVK jargon -- simulated signals to real or simulated detector noise. This functionality will be explored in the next section to predict constraints on BH charges obtainable with future detectors upgrades. The \texttt{pyRing} package has been used to produce the first ringdown-only catalog of remnant properties, together with constraints on deviations from GR QNM spectra, using data from the first three observing runs of the LIGO-Virgo interferometers; see Tables VIII-IX of Ref.~\cite{O3a_TGR}. Moreover, it has been employed to explore possible signatures~\cite{FOIT-KLEBAN, CARDOSO-FOIT-KLEBAN, AGULLO2020} of the area quantisation on the BH ringdown emission in Ref.~\cite{Laghi:2020rgl} and to obtain bounds~\cite{Carullo:2021dui} on a possible new physics length scale entering QNM spectra, in a linearized perturbative scheme~\cite{ParSpec}.

\textit{GW model --} To construct our model for a charged BH, we start from a standard $\mathrm{Kerr}$ template~\cite{Berti_fits, Lim:2019xrb}:

\be\label{eq:Kerr_model}
h_+ - i h_{\times} = \frac{M_f}{D_L} \sum_{\ell=2}^{\infty} \sum_{m=-\ell}^{+\ell} \sum_{n=0}^{\infty}\, \, (h^{+}_{\ell m n} + h^{-}_{\ell m n})
\ee

with:

\begin{subequations}
\begin{gather}
h^{+}_{\ell m n} = \mathcal{A}^{+}_{\ell m n} \, S_{\ell m n}( \iota, \varphi) \, e^{-i(t-t_{\ell m n})\tilde{\omega}_{\ell m n}+i\phi^{+}_{\ell m n}} \\
h^{-}_{\ell m n} = \mathcal{A}^{-}_{\ell m n} \, S^{*}_{\ell m n}(\pi-\iota, \varphi) \, e^{+i(t-t_{\ell m n})\tilde{\omega}^*_{\ell m n}+i\phi^{-}_{\ell m n}} 
\end{gather}
\end{subequations}
where $\tilde{\omega}_{\ell m n} = {\omega}_{\ell m n} - i/{\tau_{\ell m n}}$  (a * denotes complex conjugation) is the complex ringdown frequency, determined in the Kerr cases by the remnant BH mass $M_f$ and spin $\chi_{\scriptscriptstyle f}$,\footnote{The ``f" subscript on BH parameters indicate these values refer those of the remnant BH.} $\tilde{\omega}_{\ell m n} = \tilde{\omega}_{\ell m n}(M_f, \chi_{\scriptscriptstyle f})$.
The amplitudes $\mathcal{A}^{+/-}_{\ell m n}$ and phases $\phi^{+/-}_{\ell m n}$ characterise the excitation of each mode and are inferred from the data. The inclination of the BH final spin relative to the observer's line of sight is denoted by $\iota$, while $\varphi$ corresponds to the azimuthal angle of the line of sight in the BH frame, which without loss of generality we set to zero given the complete degeneracy with the single-mode phases. 
$S_{\ell m n}$ are the spin-weighted spheroidal harmonics~\cite{Berti:2014fga} and $t_{\ell m n}=t_0$ is a reference start time.
In writing Eq.~(\ref{eq:Kerr_model}), we follow the convention of Ref.~\cite{Lim:2019xrb} (see their Section III), for which $m>0$ indices denote co-rotating modes, while counter-rotating modes are labeled by $m<0$. In the remainder of this work, we will only consider co-rotating modes, since counter-rotating modes are predicted to be hardly excited in the post-merger phase for the binaries analysed in this work. For a discussion about the possible relevance of counter-rotating modes see Refs.~\cite{Lim:2019xrb, Dhani:2020nik, Dhani:2021vac}.

We restrict this template to a superposition of the quadrupolar fundamental (longest-lived) mode and its first corresponding overtone ($\ell=m=2, n=0,1$), considering all the amplitudes and phases as independent numbers. We refer to the template constructed in this manner using the Kerr QNM frequencies as $\mathrm{Kerr_{221}}$~\cite{Giesler:2019uxc, ISI-GIESLER-FARR-SCHEEL-TEUKOLSKY, O3a_TGR}.
The template is then modified by replacing Kerr QNM complex frequencies as a function of the remnant mass and spin $\tilde{\omega}_{\ell m n}(M_f, \chi_{\scriptscriptstyle f})$, with the corresponding KN frequencies $\tilde{\omega}_{\ell m n} = \tilde{\omega}_{\ell m n}(M_f, \chi_{\scriptscriptstyle f}, \bar{q}_{\scriptscriptstyle f})$. In the following applications, we interpolate the numerical values obtained in Sec.~\ref{sec:QNM_NR}. This modified template, used in the remainder of this work, is labeled $\mathrm{KN_{221}}$.
The assumption lying behind the construction of our template is that the post-merger signal of a BBH coalescence giving rise to a KN BH can be described by the superposition of the fundamental QNM and its first overtone. We stress that the amplitudes and phases of the modes considered in this model do not assume the predictions for a Kerr BH arising from a BBH coalescence, a key ingredient to avoid biases in the remnant PE in alternative scenarios.
Due to the high flexibility of our template, our modeling hypothesis appears robust. Nevertheless, in the future it would be interesting to directly test this assumption by comparing to numerical simulations~\cite{Bozzola:2020mjx, Bozzola:2021elc}, which would also allow to predict the values of the post-merger amplitudes and phases as a function of the binary parameters, improving the sensitivity of the model to charge effects.
Due to the coupling of EM fields to the gravitational field, in principle also the coupling of the $s=-1$ modes to the GW spectrum should be considered. However, as shown in Ref.~\cite{2016JCAP...05..054C} for simplified settings, the contribution to the gravitational emission of these modes is subdominant for non-extremal cases. Thus, we will neglect the contribution of such modes, leaving investigations of their contribution to the GW signal to future work.

\textit{Analysis details --} 
The event selection criteria (a positive Bayes factor for the hypothesis of a signal being present in the data compared to the noise-only hypothesis, and informative parameters distributions), strain data, data conditioning methods, and sampler settings are chosen to be identical to those of Ref.~\cite{O3a_TGR}, which are publicly available from the accompanying data release~\cite{O3a_TGR_data_release}. 
Additionally, for completeness we include in our analysis GW170729~\cite{Chatziioannou:2019dsz}, which was included in the testing GR analyses of the first LVC catalog~\cite{LIGOScientific:2019fpa}, but did not pass the stricter threshold imposed for the testing GR analysis of the later GWTC-2 catalog~\cite{O3a_TGR}. 
The dataset thus consists of 18 BBH events listed in Table~\ref{tab:O1O2O3_events} (out of a total of 46, mainly due to the limited sensitivity of GW detectors to high frequencies) detected by the LVC.\footnote{The selection criterion should in principle be revised in light of the new physics present in our model. 
Nevertheless, we checked that none of the excluded events passes the Bayes factor threshold applied in Ref.~\cite{O3a_TGR} or provides any significant constraint on the presence of a BH charge.}
The dataset is conservatively restricted to minimise the effect of noise events, possibly mimicking a GW event and contaminating our analysis.
The time origin of the strain for the analysis is set by the peak of $h_+^2 + h_{\times}^2$ in each of the detectors, as computed a-posteriori from an IMR analysis, and assuming the maximum likelihood value of the event sky location~\cite{O3a_TGR}. 
The adopted prior distributions are also identical to the ones chosen in Ref.~\cite{O3a_TGR}, in particular uniform on the remnant mass and spin, the latter spanning the range [0, 0.99]. The prior distribution on the charge parameter is also uniform in the interval [0, 0.99]. Finally, we impose an \emph{a priori} joint limit on the charge and spin parameters $\chi_{\scriptscriptstyle f}^2 + \bar{q}_{\scriptscriptstyle f}^2 < 0.99$, excluding near-extremal BHs configurations consistently with the numerical fits discussed in Sec.~\ref{sec:QNM_fits}.

\subsection{Analysis of the GW transient catalog}

\begin{figure*}[!tb]
\includegraphics[width=0.48\textwidth]{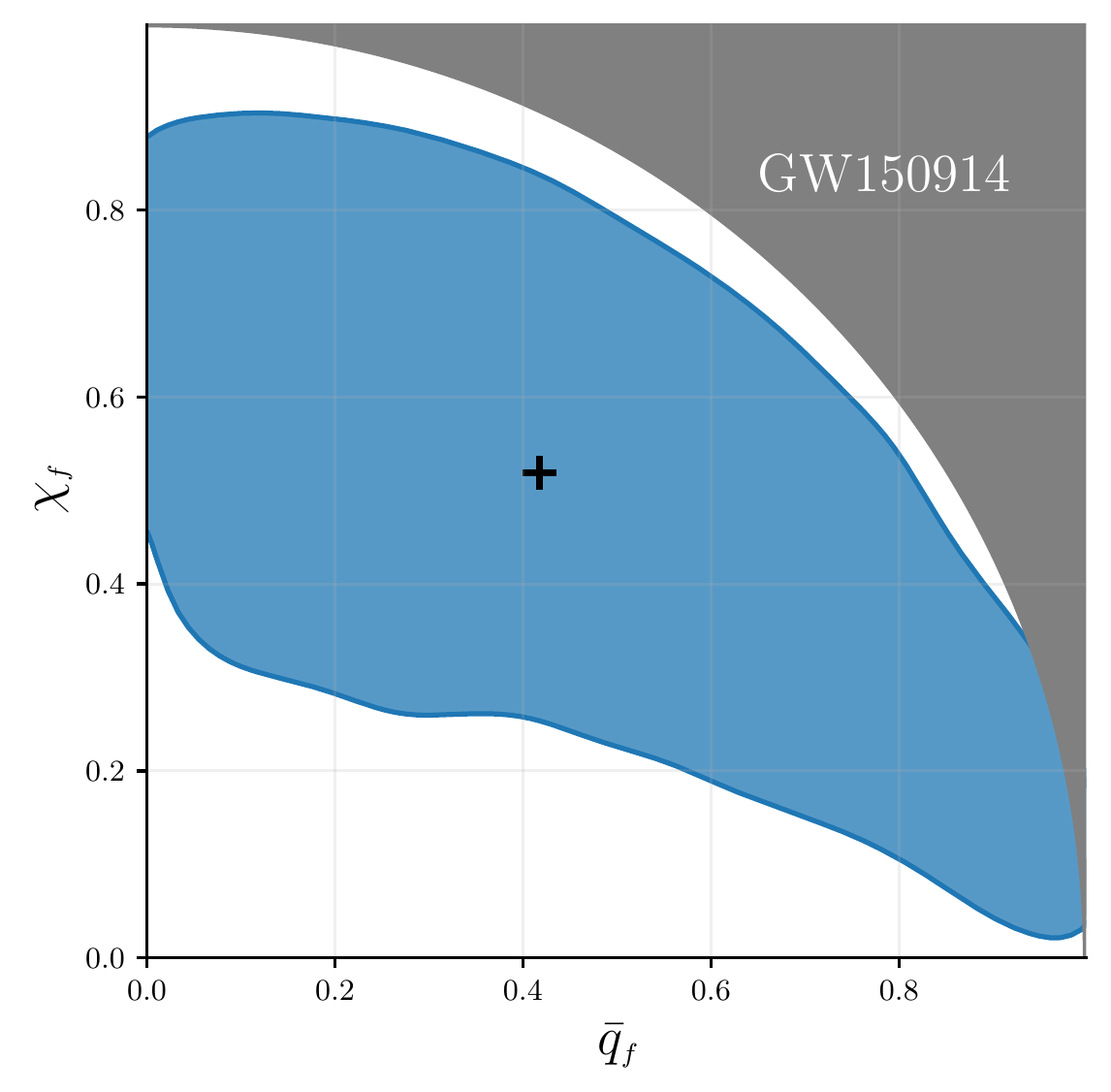}
\includegraphics[width=0.48\textwidth]{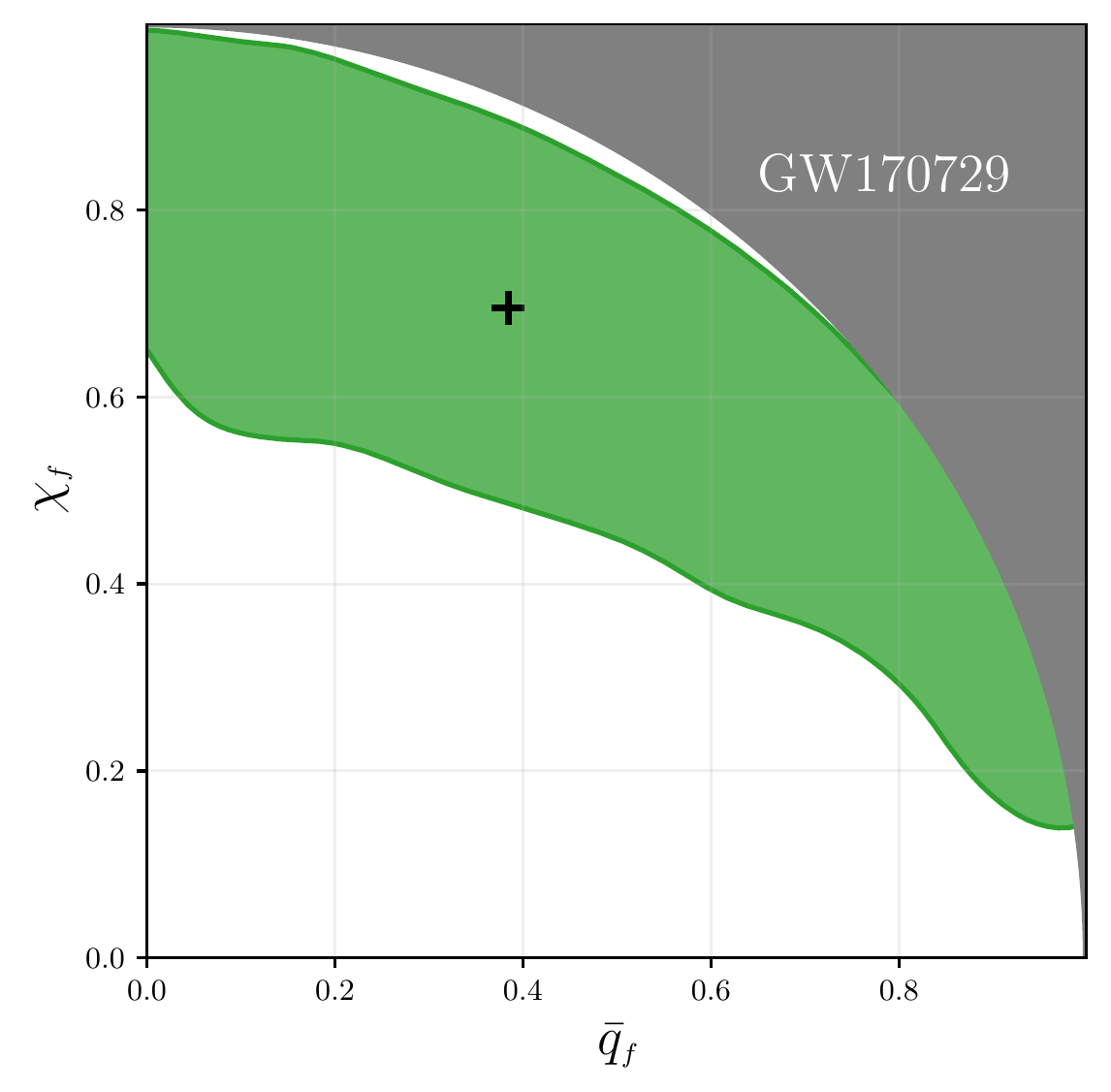}\\
\vspace{0.3cm}
\includegraphics[width=0.48\textwidth]{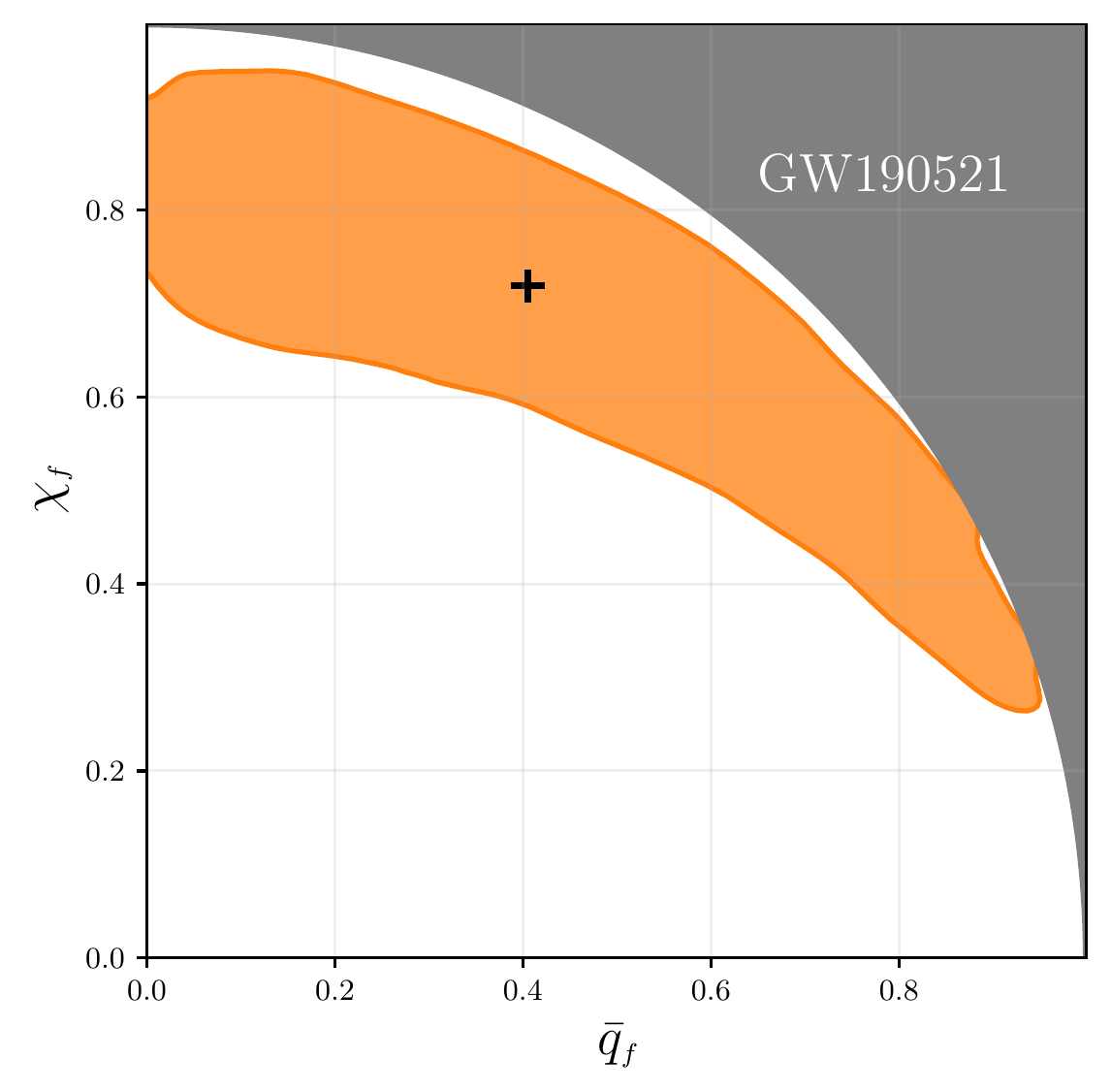}
\includegraphics[width=0.48\textwidth]{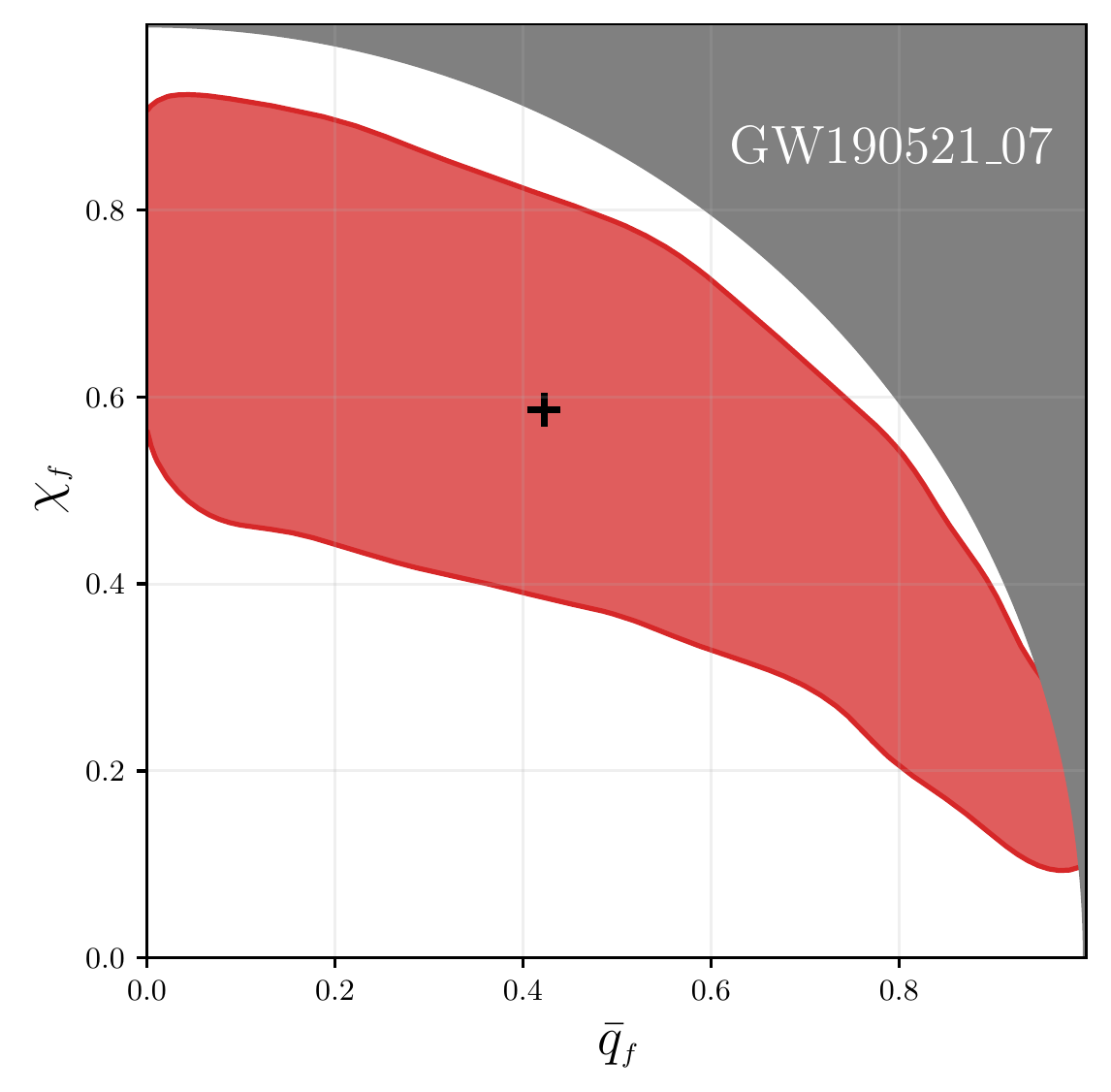}
\caption{Credible region ($90 \%$ confidence) of the two-dimensional posterior probability density function of the spin and charge-to-mass ratio, for the subset of GWTC-2 events showing the tightest constraints (GW190521\_07 stands for GW190521\_074359). Crosses mark the median of the two-dimensional distribution. The gray region marks charge-spin values above the extremal limit, excluded in the analysis. Most of the two-dimensional plane is excluded by the data, although the strong correlation between the two parameters results in a iso-frequency contour, with one-dimensional projections extending over most of the charge and spin ranges.}
\label{fig:GWTC-2_results}
\end{figure*}

\textit{Full analysis --} We apply the KN waveform model described above to the available LIGO-Virgo events selected in the previous section.
The results are presented in Fig.~\ref{fig:GWTC-2_results}, where we show the $90\%$ CL of the two-dimensional posterior distributions on remnant spin and charge-to-mass ratio, for a representative set of four events showing the strongest constraints on these parameters.
Uni-dimensional posteriors on the charge-to-mass ratio are uninformative, while the ones on the spin parameter are consistent with the result from the $\mathrm{Kerr}_{221}$ analysis, with a corresponding broadening due to the increased number of parameters included in the analysis presented here.
Remnant masses, showing very weak correlations with the charge, are always consistent with the values inferred without assuming the presence of a charge~\cite{O3a_TGR}, with a broadening analogous to the one of the spin.
Current events allow us to exclude a large portion of the spin-charge parameter space, although a strong correlation is present, due to the similar effect those two parameters have on increasing QNM frequencies.
In fact, the 90$\%$ contour roughly corresponds to an iso-frequency region, containing the inferred values of spin and charge-to-mass ratio needed to reproduce the dominant -- slowly evolving -- frequency content observed in the post-merger signal.
The typical value of the remnant spin $\chi_{\scriptscriptstyle f}$ generated by the coalescence of close-to-equal-mass, mildly spinning BHs on a quasi-circular orbit, when assuming the absence of charge, is $\chi_{\scriptscriptstyle f} \sim 0.7$~\cite{PhysRevD.77.026004, Baker:2008mj}. Around this value, the results show consistency with $\bar{q}_{\scriptscriptstyle f} \sim 0$, although the wide distribution does not allow us to strongly constrain a specific value of normalised charge and spin.
We compute a global figure of merit comparing the Kerr and KN hypotheses against the GW data, the Bayes factor. In the Kerr case we assume the same template, but now using Kerr predictions for the QNM spectra as a function of the remnant parameters. The results are reported in the first column of Table~\ref{tab:O1O2O3_events}, indicating that current data do not allow us to meaningfully distinguish between the two hypotheses within current statistical uncertainties, according to criteria such as the Jeffreys scale~\cite{Jeffreys}.

%-----------------------
\begin{table}[t]
\caption{Summary of the signal-to-noise Bayes factors between the $\mathrm{KN}$ and $\mathrm{Kerr}$ models, for both the full and null analyses, together with upper bounds at $90\%$ credibility on the remnant BH charge $\bar{q}_{\scriptscriptstyle f}$ from the null analysis. The numerical statistical error on each ln B is $\pm \, 0.1$. No significant evidence for or against charged black holes is present.}
\vspace{0.2cm}
\resizebox{1.0\columnwidth}{!}{\begin{tabular}{@{}lccc}
\hline\hline
\multicolumn{3}{c}{\qquad \qquad \qquad \qquad \qquad GWTC-2 ringdown events} \\
\hline
\hline
Event & ln B$^{\mathrm{KN}}_{\mathrm{Kerr}}$ & ln B$^{\mathrm{KN}}_{\mathrm{Kerr}}$ (null) & $\bar{q}_{\scriptscriptstyle f}$ bound at $90 \%$ (null)\\
\hline

GW150914         & $-0.6$ & $-0.7$ & $0.33$  \\
GW170104         & $-0.2$ & $-0.6$ & $0.45$  \\
GW170729         & $-0.7$ & $-0.3$ & $0.44$  \\
GW170814         & $-0.1$ & $-0.3$ & $0.45$  \\
GW170823         & $-0.1$ & $-0.6$ & $0.45$  \\
GW190408\_181802 & $-0.1$ & $-0.6$ & $0.48$  \\
GW190512\_180714 & \,\,\,\,$0.5$ & $-0.1$ & $0.56$  \\
GW190513\_205428 & $-0.5$ & $-0.8$ & $0.43$  \\
GW190519\_153544 & \,\,\,\,$0.3$ & $-0.8$ & $0.37$  \\
GW190521         & $-0.2$ & $-0.2$ & $0.47$  \\
GW190521\_074359 & $-0.2$ & $-0.8$ & $0.41$  \\
GW190602\_175927 & \,\,\,\,$0.3$ & $-0.4$ & $0.51$  \\
GW190706\_222641 & $-0.2$ & $-0.8$ & $0.43$  \\
GW190708\_232457 & \,\,\,\,$0.2$ & $-0.4$ & $0.54$  \\
GW190727\_060333 & $-0.2$ & $-0.5$ & $0.51$  \\
GW190828\_063405 & $-0.3$ & $-0.3$ & $0.47$  \\
GW190910\_112807 & \,\,\,\,$0.1$ & $-0.6$ & $0.43$  \\
GW190915\_235702 & $-0.7$ & $-0.9$ & $0.47$  \\

\hline\hline
\end{tabular}}
\label{tab:O1O2O3_events}
\end{table}
%-----------------------

\begin{figure}[!tb]
\centering
\includegraphics[width=1.0\columnwidth]{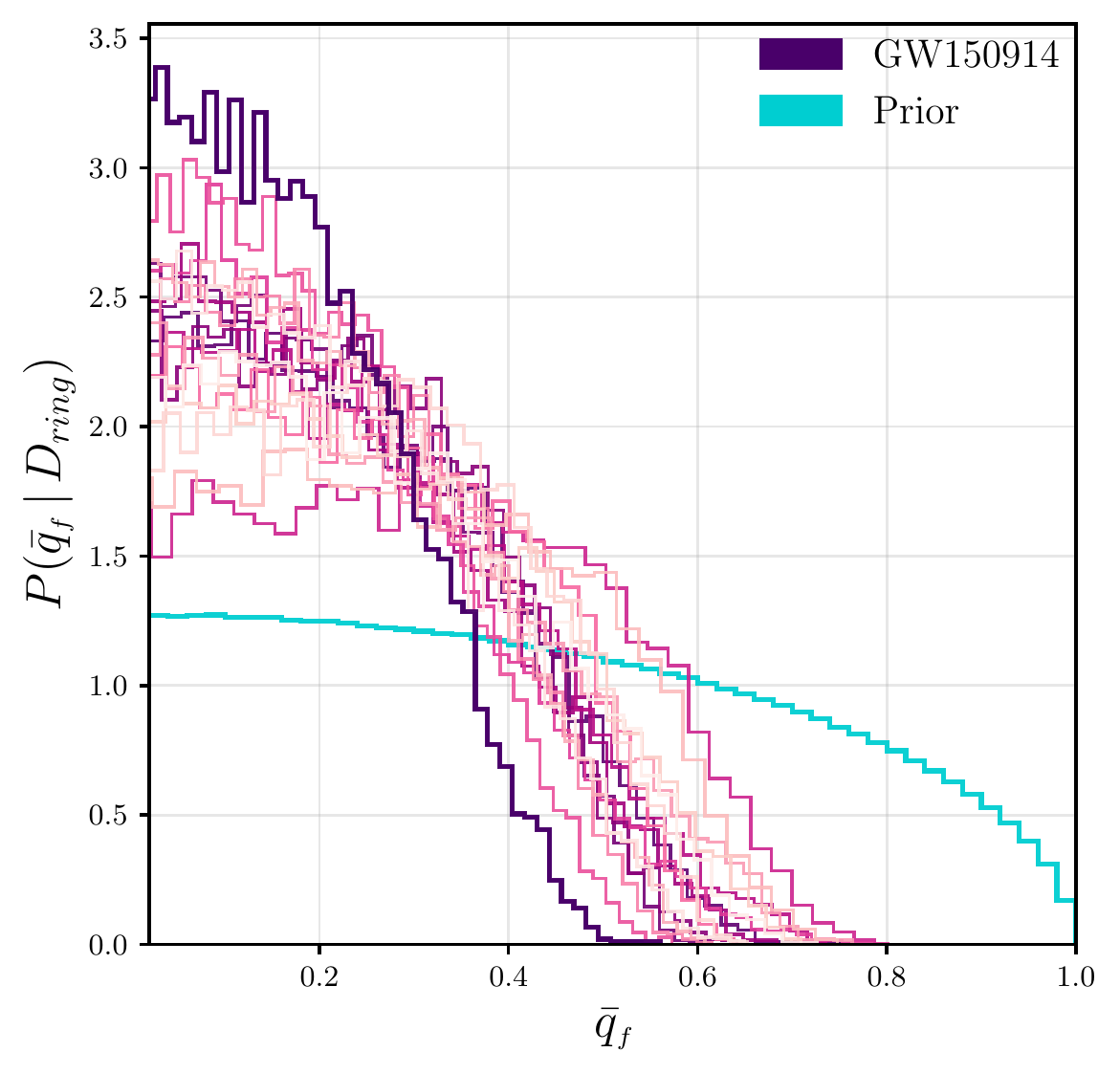}
\caption{Posterior distribution  on the charge-to-mass ratio for the null test with GWTC-2 events with detectable ringdown. The distribution is obtained from a null analysis, breaking the full correlation structure of the problem. We highlight the event showing the tightest constraint (GW150914) and the prior distribution on the charge, which incorporates the sub-extremality condition.}
\label{fig:GWTC-2_res_results}
\end{figure}

\textit{Null analysis --} As a null test, we repeat the analysis described above restricting the mass and spin uniform prior values to the $90\%$ CL obtained by the LVC collaboration from a full IMR analysis~\cite{O3a_catalog}, hence restricting them to around O($10-20\%$) of their median measured value. The outcome of such analysis will be an upper bound on the maximum allowed amount of charge compatible with LIGO-Virgo observations.
Such a test provides a comparison of our analyses with the ones discussed in the literature when ignoring the correlation of the charge with the remnant spin~\cite{Bozzola:2020mjx}.
Indeed, by restricting the available parameter space, we neglect the full correlation structure of the problem. Consequently, in the presence of an actual violation of the Kerr hypothesis, the parameter estimation resulting from this analysis could not be interpreted as the correct value of the BH charge. Nevertheless, this sort of analysis can still be used to \textit{detect} a violation of the Kerr hypothesis. In fact, if the Kerr metric is a correct description of the BH remnant, the result would yield charge values consistent with zero. By increasing the amount of information present in our inference model, this test acquires an increased accuracy on the detection of a Kerr violation, compared to the full analysis.
Results on the charge-to-mass ratio obtained under these assumptions are presented in Fig.~\ref{fig:GWTC-2_res_results}. 
In this case, the $\bar{q}_{\scriptscriptstyle f}$ posterior support is significantly reduced with respect to its prior range, the latter taking into account the sub-extremality condition $\chi_{\scriptscriptstyle f}^2 + \bar{q}_{\scriptscriptstyle f}^2 < 0.99$. 
For the most favorable case of GW150914 (highlighted in the figure), we obtain an upper bound of $\bar{q}_{\scriptscriptstyle f} < 0.33$ at 90\% credibility, consistent with the analysis presented in Ref.~\cite{Bozzola:2020mjx}. Upper bounds for the other events are reported in the rightmost column of Table~\ref{tab:O1O2O3_events}.
We recompute the Bayes factors against a Kerr hypothesis where the $M_f, \chi_{\scriptscriptstyle f}$ parameters are also restricted to the same prior bounds. 
The results are shown in the central column of Table~\ref{tab:O1O2O3_events}, again indicating that no significant evidence is present in the data for or against the KN hypothesis, as compared to the Kerr hypothesis.

\section{Future measurement prospects} 
\label{sec:Injections}

\begin{figure*}
\includegraphics[width=0.95\textwidth]{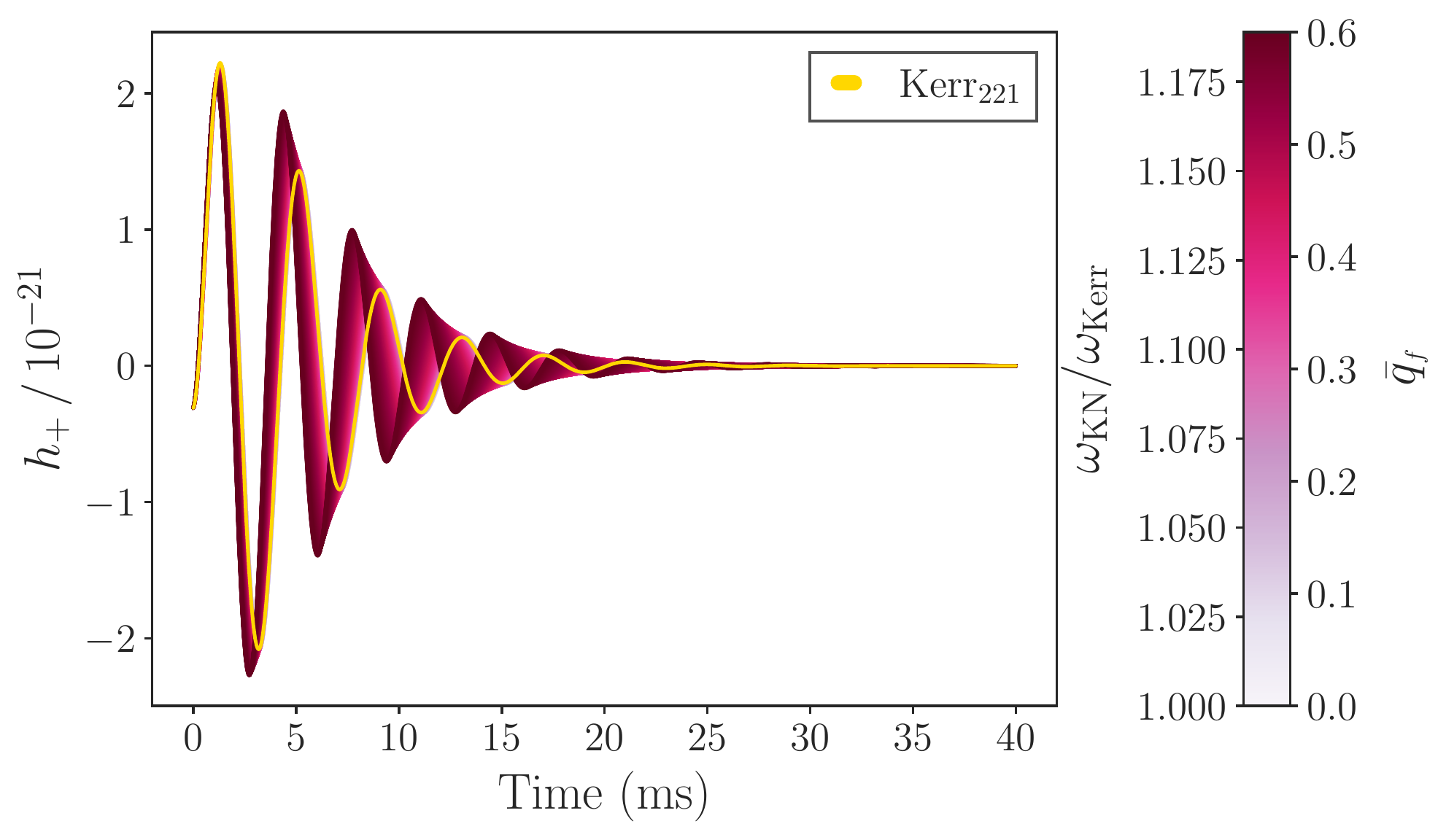}
\caption{Plus polarisation of the post-merger KN$_{221}$ model corresponding to the parameters reported in Table~\ref{tab:injection_parameters} for different values of the charge up to $\bar{q}_{\scriptscriptstyle f}=0.6$ (chosen for visualisation purposes), together with its uncharged limit, Kerr$_{221}$. The color scale is set by the charge or equivalently by the ratio of the corresponding $(\ell,m,n)=(2,2,0)$ frequencies.}
\label{fig:Waveform_KN}
\end{figure*}

Given the limited information that can be extracted from current observations, it is natural to ask whether the LIGO-Virgo network at its design sensitivity can allow us to distinguish a KN BH from a Kerr BH, using the templates considered in this work. We explore this question by addressing both the charge measurability when assuming a charged BH remnant and the sensitivity of current ringdown tests of GR when assuming uncharged BHs.

\subsection{Charge measurability} 

To address the value of charges that can be measured by the current GW detector network, we simulate charged ringdown signals, using the KN$_{221}$ template,\footnote{Our model neglects the presence of additional overtones, which we expect to be subdominant compared to the amplitude corrections induced by charged progenitors.} with increasing charge-to-mass ratio $\bar{q}_{\scriptscriptstyle f}=\{ 10^{-4}, 10^{-3}, 10^{-2}, 10^{-1}, 3 \cdot 10^{-1}, 5 \cdot 10^{-1}\}$, while the rest of the BH parameters are fixed to fiducial values close to the ones estimated for GW150914, listed in Table~\ref{tab:injection_parameters}. To reduce the number of free parameters, in our set of injections we impose the conjugate symmetry, see Ref.~\cite{Berti_fits}, $\mathcal{A}^{-}_{\ell m n} =  (-1)^l \mathcal{A}^{+}_{\ell m n}, \phi^{-}_{\ell m n} = - \phi^{+}_{\ell m n}$. The values of the amplitudes and phases are chosen from the corresponding uncharged case, by fitting the post-merger waveform of an BBH coalescence with the same intrinsic parameters, generated using the TEOBResumS model~\cite{Nagar_fits}.  We note that to obtain a good agreement between the waveforms, it is necessary to choose the relative phase $\Delta\phi$ of the fundamental mode and first overtone to be in opposition, $\Delta\phi \simeq \pi$. 
Such a requirement can be deduced from the fact that extrapolating the fundamental mode (whose amplitude is fixed by the late-time signal) back to the peak of the waveform, the corresponding peak amplitude exceeds that of a BBH remnant. 
The additional modes thus have to be chosen in such a way that the total amplitude is reduced.

Our ringdown-only reference signal has a signal-to-noise ratio (SNR) of $36$, computed by assuming the design sensitivity power spectral density of the LIGO-Virgo detector network~\cite{Aasi:2013wya}.
Fig.~\ref{fig:Waveform_KN} shows the $h_+$ polarisation corresponding to the KN$_{221}$ template for different values of charge, represented by the color scale, and the parameters reported in Table~\ref{tab:injection_parameters}. 
For each value of the charge we also indicate the ratio of the KN and Kerr fundamental frequencies. 
For a given BH spin and moderate values of the BH charge, Fig.~\ref{fig:Waveform_KN} shows a KN waveform morphology similar to Kerr, suggesting that a high SNR might be needed to distinguish the modulation of the signal due to the presence of the charge (apart from more extreme cases). The differences between the two models are further blurred by the strong correlation, which we fully account for in the analysis, between the BH spin and charge.

\begin{table*}[!tb]
\centering
\caption{
BH parameters of the KN BH ringdown signals employed in the simulation study. The table reports the injected values of final mass $M_f$, final spin $\chi_{\scriptscriptstyle f}$, real amplitudes $\mathcal{A}_{\ell mn}$ and phases $\phi_{\ell mn}$, cosine of the inclination of the BH final spin relative to the line of sight $\cos \iota$, global phase $\phi$, polarization angle $\psi$, luminosity distance $D_L$, right ascension $\alpha$, declination $\delta$ and the resulting signal-to-noise ratio, when assuming the LIGO-Virgo design sensitivity noise power spectrum.
\label{tab:injection_parameters}
}
\vspace{0.2cm}
\begin{tabular}{ccccccccccccc} 
\hline
\hline
 \multicolumn{13}{c}{Injected values} \\
\hline
\hline
$M_f \,(M_{\odot})$ & \,\,\,\, $\chi_{\scriptscriptstyle f}$ & \,\,\,\, $\mathcal{A}_{220}$ & \,\,\,\, $\mathcal{A}_{221}$ & \,\,\,\, $\phi_{220}$ & \,\,\,\, $\phi_{221}$ & \,\,\,\, $\cos \iota$ & \,\,\,\, $\phi$ & \,\,\,\, $\psi$ & \,\,\,\, $D_L \,(\text{Mpc})$ & \,\,\,\, $\alpha$ & \,\,\,\, $\delta$ & \,\,\,\, SNR \\
\hline
\hline
67.0 & \,\,\,\, 0.67 & \,\,\,\, 1.1 & \,\,\,\, 0.95 & \,\,\,\, -2.0 & \,\,\,\, 1.14 & \,\,\,\, 1.0 & \,\,\,\, 0.0 & \,\,\,\, 1.12 & \,\,\,\, 403 & \,\,\,\, 1.16 & \,\,\,\, -1.19 & \,\,\,\, 36 \\
\end{tabular}
\end{table*}

\begin{figure*}[!tb]
\includegraphics[width=0.95\textwidth]{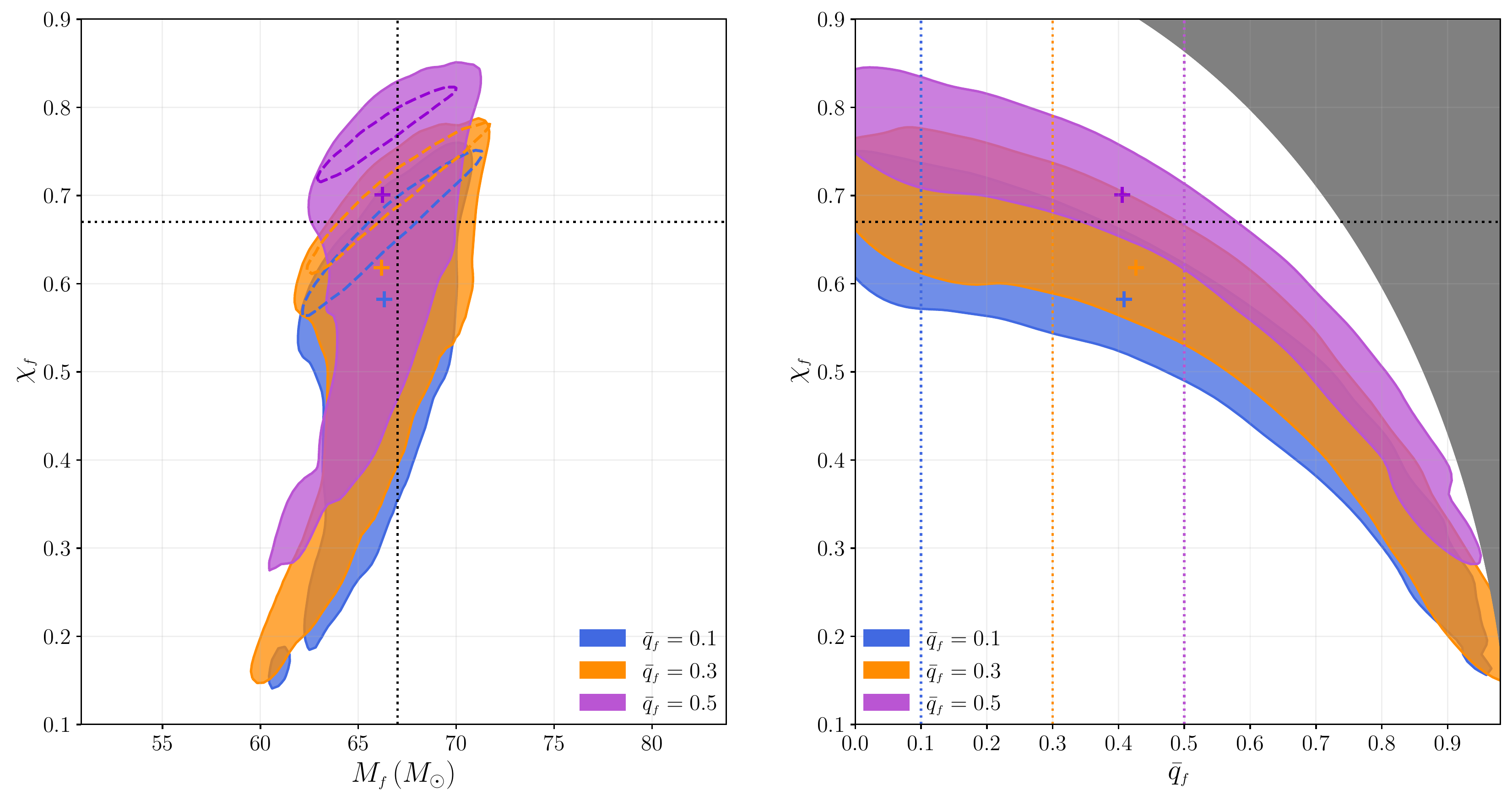}
\caption{Mass-spin and charge-spin plots for analyses of our KN$_{221}$ injections (see main text for details). Solid filled and dashed contours are KN$_{221}$ and Kerr$_{221}$ posteriors (90\% CL), respectively, with plus symbols representing the median values of the KN$_{221}$ posteriors. Dotted lines represent the injected values. The grey region in the right panel is excluded by the sub-extremality condition.}
\label{fig:injections_Mf_chif_qbar_chif_posteriors}
\end{figure*}

We perform injections of the KN templates into zero-noise, while the computation of the likelihood includes the LIGO-Virgo design sensitivity curves~\cite{Aasi:2013wya}.
This procedure is commonly adopted in the study of new physical effects in simulated LIGO-Virgo data to avoid shifts in the posteriors due to a specific noise realization.
Each of these simulated events is then recovered with different templates, corresponding to the charge and uncharged assumptions: KN$_{221}$ and Kerr$_{221}$. The first template reduces to the second in the limit of $\bar{q}_{\scriptscriptstyle f}=0$.
Analysing a KN signal with a Kerr template has the purpose of understanding the bias we would incur when ignoring a priori the presence of the BH charge, as in standard GW observational analyses.
In fact, we expect to recover a biased value of the BH spin for injections with sufficiently large values of $\bar{q}_{\scriptscriptstyle f}$, given its strong correlation with the charge parameter, as observed in the previous analysis.
This effect is illustrated in Fig.~\ref{fig:injections_Mf_chif_qbar_chif_posteriors}, where in the left panel we report posteriors (90\% CL) for mass and spin, inferred assuming the KN$_{221}$ (solid filled) and Kerr$_{221}$ (dashed line) templates for different injected values of $\bar{q}_{\scriptscriptstyle f}$, while the right panel illustrates the posteriors (90\% CL) for the charge-to-mass ratio and spin assuming the KN$_{221}$ template. Results for injections with $\bar{q}_{\scriptscriptstyle f}$ below 0.1 are very similar, so they are not shown. 

Concerning the inference with the KN$_{221}$ template, we find that the one-dimensional (marginalised) posterior for $\bar{q}_{\scriptscriptstyle f}$ is in general uninformative even for high injected values of $\bar{q}_{\scriptscriptstyle f}$, as one can also deduce from the right panel of Fig.~\ref{fig:injections_Mf_chif_qbar_chif_posteriors}, where the 90\% CL posterior extends over the whole range of $\bar{q}_{\scriptscriptstyle f}$ in the parameter space. Interestingly though, the left panel of Fig.~\ref{fig:injections_Mf_chif_qbar_chif_posteriors} suggests that the effect of a moderately large ($\bar{q} \gtrsim$ 0.3) charge-to-mass ratio on the signal could be indirectly detected: the assumption of the Kerr$_{221}$ template, i.e., excluding the presence of charge, results in a reconstructed final spin $\chi_{\scriptscriptstyle f}$ which gets increasingly biased with the value of $\bar{q}_{\scriptscriptstyle f}$. This could potentially be detected using the IMR consistency test~\cite{IMR_consistency_test1, IMR_consistency_test2}, one of the standard tests performed by the LVC. 
However, the Bayes factors are not informative enough to prefer either of the two templates, making the unique identification of such an effect (as compared to another modification of the Kerr scenario) with a BH charge difficult to obtain.
Thus, we conclude that the strong degeneracy between spin and charge does not allow for an independent measurement of the BH charge from the LIGO-Virgo network with the model considered. A similar spin-charge degeneracy is observed in the $(\ell,m,n)=(3,3,0)$ mode, suggesting that an extension of the current model considering such a mode would not strongly affect this conclusion.

\begin{figure}[!tb]
\includegraphics[width=0.99
\columnwidth]{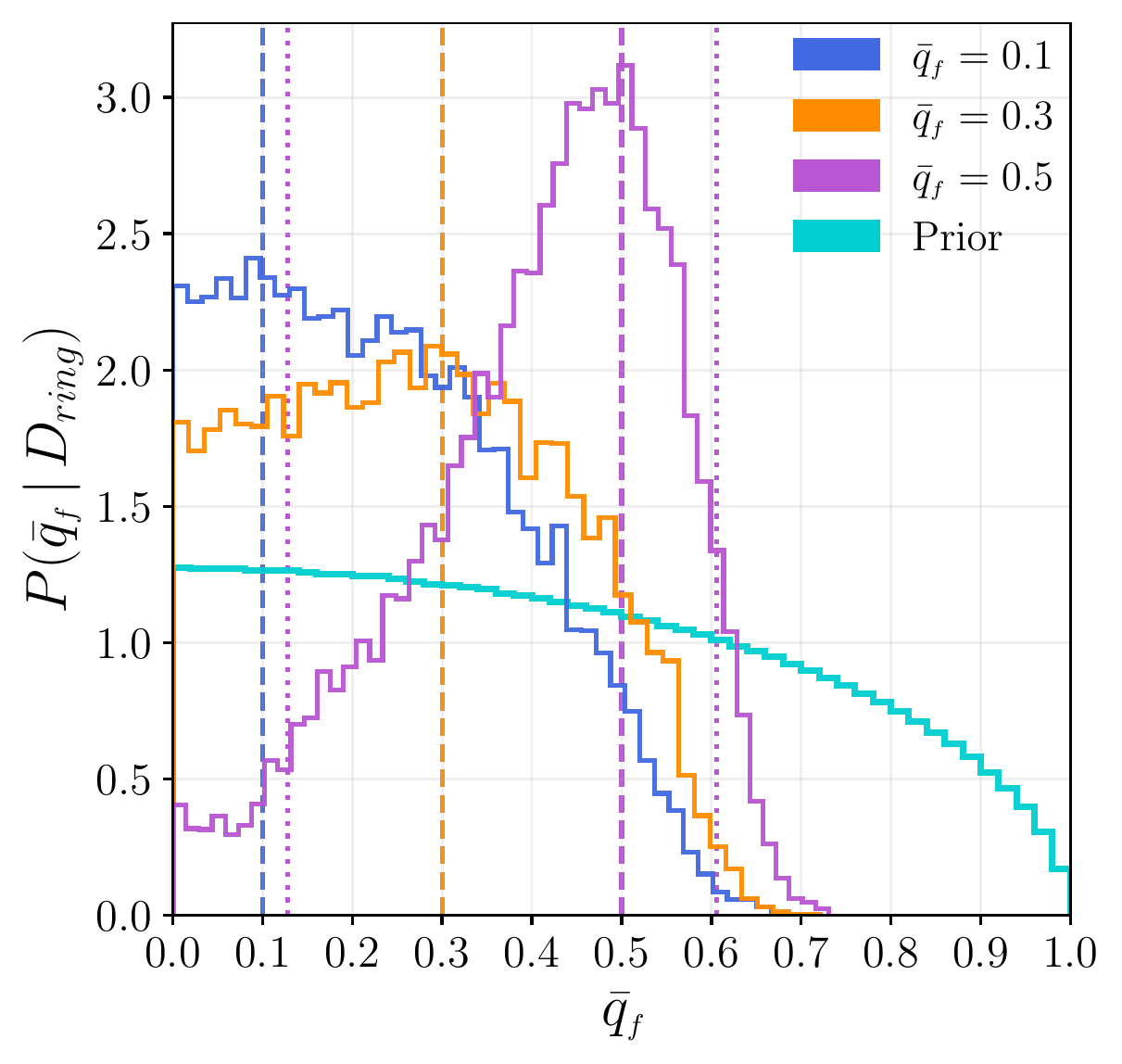}
\caption{Posterior distribution on the charge-to-mass ratio recovered analysing KN$_{221}$ signals with different values of charge-to-mass ratio (vertical dashed lines) assuming a Gaussian prior on the final spin (see the main text). For the highest charge case, we also plot the bounds on the $90\%$ CL as dotted lines.}
\label{fig:injections_charge_posteriors_gaussian_prior}
\end{figure}

As discussed in Ref.~\cite{Bozzola:2021elc}, such a correlation would instead be broken when including information from the previous stages of the coalescence, consistently modelling also the progenitors as KN BHs.
An analysis of the signal with an IMR waveform model for charged BBHs will be able to give its own estimates of the remnant charge, but an independent measurement from the ringdown will be useful to check waveform systematics and to bolster the evidence for a charged BBH detection being real and not a noise artefact.
To mimic this scenario, in Fig.~\ref{fig:injections_charge_posteriors_gaussian_prior} we report the marginalised posterior for $\bar{q}_{\scriptscriptstyle f}$ assuming a Gaussian prior constraining the final spin around its simulated value, $p(\chi_{\scriptscriptstyle f} | I) = \mathcal{N}(0.67, 0.05)$, where the width is estimated from the uncertainty associated to the 90\% CL of the final spin estimated from the IMR analysis of GW150914~\cite{LIGOScientific:2019fpa}.
The posteriors show that for $\bar{q}_{\scriptscriptstyle f} \gtrsim 0.5$ a robust measurement of the charge can be achieved, while for other values it will only be possible to place an upper bound. Our result is in agreement with the analysis of Ref.~\cite{Bozzola:2021elc}, pointing to a weak measurability up to $\bar{q}_{\scriptscriptstyle f} \sim 0.3$ at the considered SNRs.

\subsection{Tests of GR} 

Another question that naturally arises is whether the standard tests of GR routinely performed by the LVC~\cite{O3a_TGR} would signal the presence of the additional BH parameter (assuming the Kerr metric). To answer this question for the \texttt{pyRing} analysis, we consider a Kerr$_{221}$ template where now the QNM parameters are allowed to deviate with respect to the Kerr values. We consider parametric deviations of the form:
\be
    \mathrm{X} = \mathrm{X}_\text{Kerr} \cdot (1+\delta \mathrm{X}) \,, 
\ee
where $\mathrm{X}$ = $\omega_{221},\tau_{221}$. We only consider deviations in the overtone to reduce the strong degeneracy between deviations and intrinsic parameters of the BH. The fundamental mode, generally better constrained, determines the mass and spin values, while the overtone degrees of freedom are employed to constrain the deviation parameters. This allows for a much less prior-dependent determination of the deviation parameters~\cite{O3a_TGR,ghosh2021constraints}.
We define three different modified Kerr$_{221}$ templates by adding parametrised deviations either to the frequency $\{  \delta \omega_{221} \}$ or the damping time $\{ \delta \tau_{221}\}$, or to both simultaneously $\{ \delta \omega_{221}, \delta \tau_{221} \}$.
Deviations on the frequency peak around the null value for all the injected values of $\bar{q}_{\scriptscriptstyle f}$ considered, and thus do not signal any deviation from the Kerr scenario. Instead, for the highest $\bar{q}_{\scriptscriptstyle f}$ values considered, deviations on the damping times tend to be overestimated compared to the Kerr value, albeit the Kerr case is always inside the 90\% CL, thus making the test not conclusive. Similarly, the Bayes factors are uninformative, not allowing one to discriminate between templates with or without deviation parameters.
We defer further investigations to future work, possibly using more information from previous stages of the coalescence, since this should help increase the sensitivity of the test and hence its conclusiveness.

\section{Conclusions}
\label{sec:Conclusions}

In this work, we discussed extensive computations of the QNM spectrum of a KN BH for the $(\ell,m,n)=\{(2,2,0), (2,2,1), (3,3,0)\}$ modes, obtained in a companion paper~\cite{QNMsKN}, characterising the spectrum's dependence on arbitrary values of the BH charge and spin. These results were used to construct the first analytical fits of KN QNM frequencies for arbitrary values of the BH charge and spin. By extrapolating known results for the Kerr metric, we then constructed an analytical template to model the post-merger emission process of a BBH merger giving rise to a KN remnant.
We applied this model to all available LIGO-Virgo observations, showing that current data do not allow for a direct measurement of the BH charge from the post-merger emission, mainly due to the strong correlation of the charge with the remnant spin. A null test showed that the maximum value of the charge-to-mass ratio compatible with current LIGO-Virgo observations, for the most favorable event GW150914, is $\bar{q}_{\scriptscriptstyle f} < 0.33$. This is the first self-consistent observational analysis of charged remnant BHs with GWs, employing a robust statistical framework and taking the full correlation structure of the problem into account.

Finally, we performed a study aimed at exploring the sensitivity of current detectors to the remnant BH charge, finding that unless information from previous stages of the coalescence is introduced in the template, the LIGO-Virgo network at its design sensitivity will be unable to measure the charge of the remnant BH from a post-merger analysis alone. Also, current tests of GR using only the ringdown emission, routinely performed by the LVK collaboration, are unable to confidently point to a deviation from the Kerr hypothesis. However, for sufficiently large charges ($\bar{q} \gtrsim$ 0.3) a consistent overestimation of the remnant damping time (with respect to the Kerr value) could signal the presence of BH charges within the IMR consistency test.

Our results have implications for tests of general relativity and beyond Standard Model physics, since charge observations constrain the presence of magnetic monopoles, models of minicharged dark matter and alternative theories of gravity predicting the presence of an additional BH charge (through either a topological coupling or the presence of additional gravitational vector fields) degenerate with the electric charge at the scales of BH mergers.
In the future, the recent availability of full IMR simulations of charged BBHs~\cite{Bozzola:2020mjx, Bozzola:2021elc} could allow us to characterise the accuracy of the present template without relying on extrapolations of the known Kerr behaviour. Our work also provides one of the required elements for the construction of analytical templates able to model the complete signal coming from a charged BBH merger, along with the aforementioned numerical simulations and the post-Newtonian calculations in Refs.~\cite{Khalil:2018aaj,Julie:2018lfp}.

\begin{acknowledgments}

\textit{Acknowledgments}
The authors would like to thank John Veitch for useful discussions, together with Arnab Dhani and Stanislav Babak for helpful comments on the manuscript. This work greatly benefited from discussions within the \textit{Testing GR} working group of the LIGO-Virgo-KAGRA collaboration.
We gratefully acknowledge computational resources provided by the US National Science Foundation.
The authors also acknowledge the use of the cluster `Baltasar-Sete-S\'ois', and associated support services at CENTRA/IST, in the completion of this work. The authors further acknowledge the use of the IRIDIS High Performance Computing Facility, and associated support services at the University of Southampton, in the completion of this work.
This research has made use of data, software and/or web tools obtained from the Gravitational Wave Open Science Center (https://www.gw-openscience.org), a service of LIGO Laboratory, the LIGO Scientific Collaboration and the Virgo Collaboration. LIGO is funded by the U.S. National Science Foundation. Virgo is funded by the French Centre National de Recherche Scientifique (CNRS), the Italian Istituto Nazionale della Fisica Nucleare (INFN) and the Dutch Nikhef, with contributions by Polish and Hungarian institutes.
N.~K.~J.-M.\ acknowledges support from STFC Consolidator Grant No.~ST/L000636/1.
O.~C.~D.\ acknowledges financial support from the STFC ``Particle Physics Grants Panel (PPGP) 2016" Grant No.~ST/P000711/1 and the STFC ``Particle Physics Grants Panel (PPGP) 2018" Grant No.~ST/T000775/1. M.~G.\ is supported by a Royal Society University Research Fellowship. J.~E.~S.\ has been partially supported by STFC consolidated grants ST/P000681/1, ST/T000694/1. The research leading to these results has received funding from the European Research Council under the European Community's Seventh Framework Programme (FP7/2007-2013) / ERC grant agreement no. [247252].

\textit{Software} LIGO-Virgo data are interfaced through \texttt{GWpy}~\cite{gwpy} and some of the post-processing steps are handled through \texttt{PESummary}~\cite{pesummary}, a sampler-agnostic \texttt{python} package providing interactive webpages to simplify results visualisation. Moreover, \texttt{PESummary} meta-files are used to store the complete information (both of the internal \texttt{pyRing} parameters and of the software environment) for the analysis to be completely reproducible. The \texttt{pyRing} package is publicly available at: \href{https://git.ligo.org/lscsoft/pyring}{https://git.ligo.org/lscsoft/pyring}, where example configuration files using the KN spectrum are also provided.
Other open-software \texttt{python} packages, accessible through \texttt{PyPi}, internally used by \texttt{pyRing} are: \texttt{corner, matplotlib, numba, numpy, seaborn}~\cite{corner, matplotlib, numba, numpy, seaborn}.
\end{acknowledgments}

\section*{Fit coefficients}\label{sec:Appendix}

In Tables~\ref{tab:coeffs_table_r}, \ref{tab:coeffs_table_i} we report the numerical coefficients obtained fitting the data presented in Sec.~\ref{sec:QNM_NR}, using the template of Eq.~(\ref{eq:Nagar_etal_ext}) with the Bayesian method described in Sec.~\ref{sec:QNM_fits}. 
Single point estimates correspond to the maximum of the posterior distribution (the same as the maximum of the likelihood, since the priors on all coefficients are uniform), which should be used in applications where a point estimate is employed. We also report the median and $90\%$ CIs of the full probability distribution.

%-----------------------
\begin{table*}[t]
\caption{Numerical results for the coefficients of the real QNM frequency, using as a template the rational expression considered in Eq.~(\ref{eq:Nagar_etal_ext}) with $N_\text{max}=3$. The first column of each mode reports the maximum of the posterior, while the second reports median and $90\%$ CL from the full probability distribution. For applications in which a single point estimate is used, the maximum of the posterior yields a more faithful representation of the numerical data. The Schwarzschild value is given by: $Y_0 =  \{0.37367168, 0.34671099, 0.59944329 \}$ for the $(\ell,m,n) = \{ (2,2,0), (2,2,1), (3,3,0) \}$ modes respectively, while b$_{0,0}=$ c$_{0,0}=1$ by definition.}
\begin{ruledtabular}
\resizebox{1.0\textwidth}{!}{\begin{tabular}{c|ccc}
 & & $\omega$ & \\
\hline \hline
$(\ell,m,n)$ & $(2,2,0)$ & $(2,2,1)$ & $(3,3,0)$ \\
\hline \hline
 & $\mathrm{max \mathcal{P}}$ \quad \quad \quad \quad \hspace{0.1cm} $\mathrm{Prob}$ & \hspace{0.1cm} $\mathrm{max \mathcal{P}}$ \quad \quad \quad \quad \hspace{0.1cm} $\mathrm{Prob}$ & \hspace{0.1cm} $\mathrm{max \mathcal{P}}$ \quad \quad \quad \quad \hspace{0.1cm} $\mathrm{Prob}$ \\
\hline \hline
b$_{0,1}$ & \,\,\,\,\,${0.537583}$ \quad \quad \,\,\,\,\,${0.541}^{+0.045}_{-0.050}$ & ${-2.918987}$ \quad \quad ${-2.918}^{+0.001}_{-0.001}$ & ${-0.311963}$ \quad \quad ${-0.299}^{+0.019}_{-0.017}$\\
b$_{0,2}$ & ${-2.990402}$ \quad \quad ${-2.997}^{+0.084}_{-0.077}$ & \,\,\,\,\,${2.866252}$ \quad \quad \,\,\,\,\,${2.865}^{+0.002}_{-0.001}$ & ${-1.457057}$ \quad \quad ${-1.478}^{+0.028}_{-0.031}$\\
b$_{0,3}$ & \,\,\,\,\,${1.503421}$ \quad \quad \,\,\,\,\,${1.507}^{+0.032}_{-0.035}$ & ${-0.944554}$ \quad \quad ${-0.944}^{+0.001}_{-0.001}$ & \,\,\,\,\,${0.825692}$ \quad \quad \,\,\,\,\,${0.834}^{+0.013}_{-0.012}$\\
b$_{1,0}$ & ${-1.899567}$ \quad \quad ${-1.895}^{+0.005}_{-0.007}$ & ${-1.850299}$ \quad \quad ${-1.853}^{+0.004}_{-0.003}$ & ${-1.928277}$ \quad \quad ${-1.926}^{+0.003}_{-0.003}$\\
b$_{1,1}$ & ${-2.128633}$ \quad \quad ${-2.143}^{+0.120}_{-0.109}$ & \,\,\,\,\,${7.321955}$ \quad \quad \,\,\,\,\,${7.320}^{+0.005}_{-0.008}$ & ${-0.026433}$ \quad \quad ${-0.060}^{+0.040}_{-0.048}$\\
b$_{1,2}$ & \,\,\,\,\,${6.626680}$ \quad \quad \,\,\,\,\,${6.649}^{+0.163}_{-0.183}$ & ${-8.783456}$ \quad \quad ${-8.775}^{+0.020}_{-0.007}$ & \,\,\,\,\,${3.139427}$ \quad \quad \,\,\,\,\,${3.190}^{+0.071}_{-0.063}$\\
b$_{1,3}$ & ${-2.903790}$ \quad \quad ${-2.914}^{+0.069}_{-0.064}$ & \,\,\,\,\,${3.292966}$ \quad \quad \,\,\,\,\,${3.288}^{+0.004}_{-0.011}$ & ${-1.484557}$ \quad \quad ${-1.504}^{+0.026}_{-0.026}$\\
b$_{2,0}$ & \,\,\,\,\,${1.015454}$ \quad \quad \,\,\,\,\,${1.009}^{+0.010}_{-0.008}$ & \,\,\,\,\,${0.944088}$ \quad \quad \,\,\,\,\,${0.948}^{+0.005}_{-0.005}$ & \,\,\,\,\,${1.044039}$ \quad \quad \,\,\,\,\,${1.041}^{+0.004}_{-0.004}$\\
b$_{2,1}$ & \,\,\,\,\,${2.147094}$ \quad \quad \,\,\,\,\,${2.162}^{+0.087}_{-0.094}$ & ${-5.584876}$ \quad \quad ${-5.583}^{+0.010}_{-0.009}$ & \,\,\,\,\,${0.545708}$ \quad \quad \,\,\,\,\,${0.575}^{+0.037}_{-0.034}$\\
b$_{2,2}$ & ${-4.672847}$ \quad \quad ${-4.692}^{+0.129}_{-0.116}$ & \,\,\,\,\,${7.675096}$ \quad \quad \,\,\,\,\,${7.666}^{+0.010}_{-0.027}$ & ${-2.188569}$ \quad \quad ${-2.229}^{+0.048}_{-0.051}$\\
b$_{2,3}$ & \,\,\,\,\,${1.891731}$ \quad \quad \,\,\,\,\,${1.900}^{+0.044}_{-0.046}$ & ${-3.039132}$ \quad \quad ${-3.035}^{+0.012}_{-0.005}$ & \,\,\,\,\,${0.940019}$ \quad \quad \,\,\,\,\,${0.956}^{+0.019}_{-0.018}$\\
b$_{3,0}$ & ${-0.111430}$ \quad \quad ${-0.109}^{+0.003}_{-0.004}$ & ${-0.088458}$ \quad \quad ${-0.089}^{+0.001}_{-0.002}$ & ${-0.112303}$ \quad \quad ${-0.111}^{+0.002}_{-0.001}$\\
b$_{3,1}$ & ${-0.581706}$ \quad \quad ${-0.585}^{+0.022}_{-0.020}$ & \,\,\,\,\,${1.198758}$ \quad \quad \,\,\,\,\,${1.198}^{+0.004}_{-0.003}$ & ${-0.226402}$ \quad \quad ${-0.234}^{+0.008}_{-0.009}$\\
b$_{3,2}$ & \,\,\,\,\,${1.021061}$ \quad \quad \,\,\,\,\,${1.025}^{+0.028}_{-0.029}$ & ${-1.973222}$ \quad \quad ${-1.971}^{+0.009}_{-0.004}$ & \,\,\,\,\,${0.482482}$ \quad \quad \,\,\,\,\,${0.493}^{+0.012}_{-0.012}$\\
b$_{3,3}$ & ${-0.414517}$ \quad \quad ${-0.416}^{+0.011}_{-0.011}$ & \,\,\,\,\,${0.838109}$ \quad \quad \,\,\,\,\,${0.837}^{+0.002}_{-0.004}$ & ${-0.204299}$ \quad \quad ${-0.209}^{+0.005}_{-0.004}$\\
\hline \hline
c$_{0,1}$ & \,\,\,\,\,${0.548651}$ \quad \quad \,\,\,\,\,${0.552}^{+0.046}_{-0.050}$ & ${-2.941138}$ \quad \quad ${-2.940}^{+0.001}_{-0.001}$ & ${-0.299153}$ \quad \quad ${-0.286}^{+0.019}_{-0.017}$\\
c$_{0,2}$ & ${-3.141145}$ \quad \quad ${-3.148}^{+0.087}_{-0.079}$ & \,\,\,\,\,${2.907859}$ \quad \quad \,\,\,\,\,${2.907}^{+0.002}_{-0.001}$ & ${-1.591595}$ \quad \quad ${-1.613}^{+0.029}_{-0.033}$\\
c$_{0,3}$ & \,\,\,\,\,${1.636377}$ \quad \quad \,\,\,\,\,${1.640}^{+0.034}_{-0.037}$ & ${-0.964407}$ \quad \quad ${-0.964}^{+0.001}_{-0.001}$ & \,\,\,\,\,${0.938987}$ \quad \quad \,\,\,\,\,${0.948}^{+0.014}_{-0.012}$\\
c$_{1,0}$ & ${-2.238461}$ \quad \quad ${-2.235}^{+0.005}_{-0.006}$ & ${-2.250169}$ \quad \quad ${-2.253}^{+0.003}_{-0.003}$ & ${-2.265230}$ \quad \quad ${-2.263}^{+0.003}_{-0.003}$\\
c$_{1,1}$ & ${-2.291933}$ \quad \quad ${-2.307}^{+0.134}_{-0.124}$ & \,\,\,\,\,${8.425183}$ \quad \quad \,\,\,\,\,${8.423}^{+0.005}_{-0.008}$ & \,\,\,\,\,${0.058508}$ \quad \quad \,\,\,\,\,${0.022}^{+0.045}_{-0.054}$\\
c$_{1,2}$ & \,\,\,\,\,${7.695570}$ \quad \quad \,\,\,\,\,${7.718}^{+0.188}_{-0.208}$ & ${-9.852886}$ \quad \quad ${-9.844}^{+0.021}_{-0.007}$ & \,\,\,\,\,${3.772084}$ \quad \quad \,\,\,\,\,${3.828}^{+0.082}_{-0.071}$\\
c$_{1,3}$ & ${-3.458474}$ \quad \quad ${-3.470}^{+0.082}_{-0.072}$ & \,\,\,\,\,${3.660289}$ \quad \quad \,\,\,\,\,${3.655}^{+0.004}_{-0.011}$ & ${-1.852247}$ \quad \quad ${-1.874}^{+0.030}_{-0.031}$\\
c$_{2,0}$ & \,\,\,\,\,${1.581677}$ \quad \quad \,\,\,\,\,${1.575}^{+0.011}_{-0.009}$ & \,\,\,\,\,${1.611393}$ \quad \quad \,\,\,\,\,${1.616}^{+0.005}_{-0.006}$ & \,\,\,\,\,${1.624332}$ \quad \quad \,\,\,\,\,${1.621}^{+0.005}_{-0.005}$\\
c$_{2,1}$ & \,\,\,\,\,${2.662938}$ \quad \quad \,\,\,\,\,${2.682}^{+0.115}_{-0.124}$ & ${-7.869432}$ \quad \quad ${-7.867}^{+0.013}_{-0.008}$ & \,\,\,\,\,${0.533096}$ \quad \quad \,\,\,\,\,${0.569}^{+0.050}_{-0.043}$\\
c$_{2,2}$ & ${-6.256090}$ \quad \quad ${-6.281}^{+0.170}_{-0.157}$ & \,\,\,\,\,${9.999751}$ \quad \quad \,\,\,\,\,${9.988}^{+0.011}_{-0.032}$ & ${-3.007197}$ \quad \quad ${-3.056}^{+0.061}_{-0.067}$\\
c$_{2,3}$ & \,\,\,\,\,${2.494264}$ \quad \quad \,\,\,\,\,${2.506}^{+0.055}_{-0.060}$ & ${-3.737205}$ \quad \quad ${-3.731}^{+0.014}_{-0.005}$ & \,\,\,\,\,${1.285026}$ \quad \quad \,\,\,\,\,${1.303}^{+0.024}_{-0.023}$\\
c$_{3,0}$ & ${-0.341455}$ \quad \quad ${-0.338}^{+0.004}_{-0.005}$ & ${-0.359285}$ \quad \quad ${-0.361}^{+0.002}_{-0.002}$ & ${-0.357651}$ \quad \quad ${-0.356}^{+0.002}_{-0.002}$\\
c$_{3,1}$ & ${-0.930069}$ \quad \quad ${-0.937}^{+0.037}_{-0.034}$ & \,\,\,\,\,${2.392321}$ \quad \quad \,\,\,\,\,${2.391}^{+0.003}_{-0.005}$ & ${-0.300599}$ \quad \quad ${-0.311}^{+0.012}_{-0.015}$\\
c$_{3,2}$ & \,\,\,\,\,${1.688288}$ \quad \quad \,\,\,\,\,${1.697}^{+0.042}_{-0.046}$ & ${-3.154979}$ \quad \quad ${-3.151}^{+0.012}_{-0.005}$ & \,\,\,\,\,${0.810387}$ \quad \quad \,\,\,\,\,${0.824}^{+0.018}_{-0.017}$\\
c$_{3,3}$ & ${-0.612643}$ \quad \quad ${-0.616}^{+0.015}_{-0.014}$ & \,\,\,\,\,${1.129776}$ \quad \quad \,\,\,\,\,${1.128}^{+0.002}_{-0.005}$ & ${-0.314715}$ \quad \quad ${-0.320}^{+0.006}_{-0.006}$\vspace{0.05cm}\\
\end{tabular}}
\end{ruledtabular}
\label{tab:coeffs_table_r}
\end{table*}
%-----------------------

%-----------------------
\begin{table*}[t]
\caption{Numerical results for the coefficients of the QNM inverse damping time, using as a template the rational expression considered in Eq.~(\ref{eq:Nagar_etal_ext}) with $N_\text{max}=3$. The first column of each mode reports the maximum of the posterior, while the second reports median and $90\%$ CL from the full probability distribution. For applications in which a single point estimate is used, the maximum of the posterior yields a more faithful representation of the numerical data. The Schwarzschild value is given by: $Y_0 =  \{0.08896232, 0.27391488, 0.09270305 \}$ for the $(\ell,m,n) = \{ (2,2,0), (2,2,1), (3,3,0) \}$ modes respectively, while b$_{0,0}=$ c$_{0,0}=1$ by definition.}
\begin{ruledtabular}
\resizebox{1.0\textwidth}{!}{\begin{tabular}{c|ccc}
 & & $\tau^{-1}$ & \\
\hline \hline
$(\ell,m,n)$ & $(2,2,0)$ & $(2,2,1)$ & $(3,3,0)$ \\
\hline \hline
 & $\mathrm{max \mathcal{P}}$ \quad \quad \quad \quad \hspace{0.1cm} $\mathrm{Prob}$ & \hspace{0.1cm} $\mathrm{max \mathcal{P}}$ \quad \quad \quad \quad \hspace{0.1cm} $\mathrm{Prob}$ & \hspace{0.1cm} $\mathrm{max \mathcal{P}}$ \quad \quad \quad \quad \hspace{0.1cm} $\mathrm{Prob}$ \\
\hline \hline
b$_{0,1}$ & ${-2.721789}$ \quad \quad ${-2.723}^{+0.016}_{-0.014}$ & ${-3.074983}$ \quad \quad ${-3.073}^{+0.005}_{-0.005}$ & ${-2.813977}$ \quad \quad ${-2.817}^{+0.018}_{-0.017}$\\
b$_{0,2}$ & \,\,\,\,\,${2.472860}$ \quad \quad \,\,\,\,\,${2.476}^{+0.028}_{-0.031}$ & \,\,\,\,\,${3.182195}$ \quad \quad \,\,\,\,\,${3.179}^{+0.009}_{-0.009}$ & \,\,\,\,\,${2.666759}$ \quad \quad \,\,\,\,\,${2.672}^{+0.033}_{-0.033}$\\
b$_{0,3}$ & ${-0.750015}$ \quad \quad ${-0.752}^{+0.015}_{-0.014}$ & ${-1.105297}$ \quad \quad ${-1.103}^{+0.005}_{-0.004}$ & ${-0.850618}$ \quad \quad ${-0.853}^{+0.016}_{-0.017}$\\
b$_{1,0}$ & ${-2.533958}$ \quad \quad ${-2.519}^{+0.024}_{-0.022}$ & \,\,\,\,\,${0.366066}$ \quad \quad \,\,\,\,\,${0.343}^{+0.046}_{-0.048}$ & ${-2.163575}$ \quad \quad ${-2.161}^{+0.035}_{-0.035}$\\
b$_{1,1}$ & \,\,\,\,\,${7.181110}$ \quad \quad \,\,\,\,\,${7.173}^{+0.062}_{-0.061}$ & \,\,\,\,\,${4.296285}$ \quad \quad \,\,\,\,\,${4.328}^{+0.067}_{-0.065}$ & \,\,\,\,\,${6.934304}$ \quad \quad \,\,\,\,\,${6.969}^{+0.095}_{-0.093}$\\
b$_{1,2}$ & ${-6.870324}$ \quad \quad ${-6.898}^{+0.099}_{-0.109}$ & ${-9.700146}$ \quad \quad ${-9.696}^{+0.011}_{-0.012}$ & ${-7.425335}$ \quad \quad ${-7.499}^{+0.147}_{-0.160}$\\
b$_{1,3}$ & \,\,\,\,\,${2.214689}$ \quad \quad \,\,\,\,\,${2.236}^{+0.053}_{-0.049}$ & \,\,\,\,\,${5.016955}$ \quad \quad \,\,\,\,\,${5.004}^{+0.026}_{-0.027}$ & \,\,\,\,\,${2.640936}$ \quad \quad \,\,\,\,\,${2.679}^{+0.077}_{-0.072}$\\
b$_{2,0}$ & \,\,\,\,\,${2.102750}$ \quad \quad \,\,\,\,\,${2.075}^{+0.043}_{-0.047}$ & ${-3.290350}$ \quad \quad ${-3.247}^{+0.091}_{-0.088}$ & \,\,\,\,\,${1.405496}$ \quad \quad \,\,\,\,\,${1.401}^{+0.068}_{-0.067}$\\
b$_{2,1}$ & ${-6.317887}$ \quad \quad ${-6.300}^{+0.092}_{-0.093}$ & ${-0.844265}$ \quad \quad ${-0.904}^{+0.119}_{-0.123}$ & ${-5.678573}$ \quad \quad ${-5.739}^{+0.149}_{-0.157}$\\
b$_{2,2}$ & \,\,\,\,\,${6.206452}$ \quad \quad \,\,\,\,\,${6.249}^{+0.126}_{-0.117}$ & \,\,\,\,\,${9.999863}$ \quad \quad \,\,\,\,\,${9.999}^{+0.001}_{-0.002}$ & \,\,\,\,\,${6.621826}$ \quad \quad \,\,\,\,\,${6.739}^{+0.226}_{-0.204}$\\
b$_{2,3}$ & ${-1.980749}$ \quad \quad ${-2.007}^{+0.052}_{-0.062}$ & ${-5.818349}$ \quad \quad ${-5.802}^{+0.034}_{-0.031}$ & ${-2.345713}$ \quad \quad ${-2.401}^{+0.092}_{-0.101}$\\
b$_{3,0}$ & ${-0.568636}$ \quad \quad ${-0.555}^{+0.022}_{-0.021}$ & \,\,\,\,\,${1.927196}$ \quad \quad \,\,\,\,\,${1.906}^{+0.041}_{-0.043}$ & ${-0.241561}$ \quad \quad ${-0.240}^{+0.033}_{-0.032}$\\
b$_{3,1}$ & \,\,\,\,\,${1.857404}$ \quad \quad \,\,\,\,\,${1.851}^{+0.040}_{-0.041}$ & ${-0.401520}$ \quad \quad ${-0.376}^{+0.054}_{-0.052}$ & \,\,\,\,\,${1.555843}$ \quad \quad \,\,\,\,\,${1.584}^{+0.072}_{-0.068}$\\
b$_{3,2}$ & ${-1.820547}$ \quad \quad ${-1.836}^{+0.047}_{-0.050}$ & ${-3.537667}$ \quad \quad ${-3.537}^{+0.003}_{-0.003}$ & ${-1.890365}$ \quad \quad ${-1.942}^{+0.085}_{-0.087}$\\
b$_{3,3}$ & \,\,\,\,\,${0.554722}$ \quad \quad \,\,\,\,\,${0.564}^{+0.021}_{-0.018}$ & \,\,\,\,\,${2.077991}$ \quad \quad \,\,\,\,\,${2.072}^{+0.012}_{-0.013}$ & \,\,\,\,\,${0.637480}$ \quad \quad \,\,\,\,\,${0.659}^{+0.035}_{-0.032}$\\
\hline \hline
c$_{0,1}$ & ${-2.732346}$ \quad \quad ${-2.734}^{+0.016}_{-0.014}$ & ${-3.079686}$ \quad \quad ${-3.078}^{+0.005}_{-0.005}$ & ${-2.820763}$ \quad \quad ${-2.823}^{+0.017}_{-0.016}$\\
c$_{0,2}$ & \,\,\,\,\,${2.495049}$ \quad \quad \,\,\,\,\,${2.498}^{+0.027}_{-0.029}$ & \,\,\,\,\,${3.191889}$ \quad \quad \,\,\,\,\,${3.188}^{+0.009}_{-0.009}$ & \,\,\,\,\,${2.680557}$ \quad \quad \,\,\,\,\,${2.686}^{+0.031}_{-0.033}$\\
c$_{0,3}$ & ${-0.761581}$ \quad \quad ${-0.763}^{+0.014}_{-0.013}$ & ${-1.110140}$ \quad \quad ${-1.108}^{+0.004}_{-0.004}$ & ${-0.857462}$ \quad \quad ${-0.860}^{+0.016}_{-0.016}$\\
c$_{1,0}$ & ${-2.498341}$ \quad \quad ${-2.484}^{+0.024}_{-0.022}$ & \,\,\,\,\,${0.388928}$ \quad \quad \,\,\,\,\,${0.366}^{+0.046}_{-0.048}$ & ${-2.130446}$ \quad \quad ${-2.128}^{+0.035}_{-0.035}$\\
c$_{1,1}$ & \,\,\,\,\,${7.089542}$ \quad \quad \,\,\,\,\,${7.080}^{+0.062}_{-0.060}$ & \,\,\,\,\,${4.159242}$ \quad \quad \,\,\,\,\,${4.192}^{+0.068}_{-0.066}$ & \,\,\,\,\,${6.825101}$ \quad \quad \,\,\,\,\,${6.858}^{+0.095}_{-0.091}$\\
c$_{1,2}$ & ${-6.781334}$ \quad \quad ${-6.807}^{+0.096}_{-0.104}$ & ${-9.474149}$ \quad \quad ${-9.472}^{+0.010}_{-0.010}$ & ${-7.291058}$ \quad \quad ${-7.361}^{+0.142}_{-0.157}$\\
c$_{1,3}$ & \,\,\,\,\,${2.181880}$ \quad \quad \,\,\,\,\,${2.201}^{+0.051}_{-0.046}$ & \,\,\,\,\,${4.904881}$ \quad \quad \,\,\,\,\,${4.893}^{+0.024}_{-0.025}$ & \,\,\,\,\,${2.583282}$ \quad \quad \,\,\,\,\,${2.619}^{+0.074}_{-0.070}$\\
c$_{2,0}$ & \,\,\,\,\,${2.056918}$ \quad \quad \,\,\,\,\,${2.030}^{+0.041}_{-0.045}$ & ${-3.119527}$ \quad \quad ${-3.077}^{+0.087}_{-0.085}$ & \,\,\,\,\,${1.394144}$ \quad \quad \,\,\,\,\,${1.390}^{+0.065}_{-0.065}$\\
c$_{2,1}$ & ${-6.149334}$ \quad \quad ${-6.132}^{+0.090}_{-0.089}$ & ${-0.914668}$ \quad \quad ${-0.974}^{+0.117}_{-0.119}$ & ${-5.533669}$ \quad \quad ${-5.589}^{+0.143}_{-0.151}$\\
c$_{2,2}$ & \,\,\,\,\,${6.010021}$ \quad \quad \,\,\,\,\,${6.048}^{+0.120}_{-0.113}$ & \,\,\,\,\,${9.767356}$ \quad \quad \,\,\,\,\,${9.768}^{+0.005}_{-0.005}$ & \,\,\,\,\,${6.393699}$ \quad \quad \,\,\,\,\,${6.504}^{+0.213}_{-0.193}$\\
c$_{2,3}$ & ${-1.909275}$ \quad \quad ${-1.933}^{+0.050}_{-0.058}$ & ${-5.690517}$ \quad \quad ${-5.676}^{+0.033}_{-0.030}$ & ${-2.254239}$ \quad \quad ${-2.306}^{+0.087}_{-0.097}$\\
c$_{3,0}$ & ${-0.557557}$ \quad \quad ${-0.545}^{+0.021}_{-0.020}$ & \,\,\,\,\,${1.746957}$ \quad \quad \,\,\,\,\,${1.728}^{+0.038}_{-0.040}$ & ${-0.261229}$ \quad \quad ${-0.260}^{+0.030}_{-0.030}$\\
c$_{3,1}$ & \,\,\,\,\,${1.786783}$ \quad \quad \,\,\,\,\,${1.780}^{+0.038}_{-0.039}$ & ${-0.240680}$ \quad \quad ${-0.216}^{+0.049}_{-0.050}$ & \,\,\,\,\,${1.517744}$ \quad \quad \,\,\,\,\,${1.543}^{+0.067}_{-0.064}$\\
c$_{3,2}$ & ${-1.734461}$ \quad \quad ${-1.749}^{+0.046}_{-0.047}$ & ${-3.505359}$ \quad \quad ${-3.505}^{+0.004}_{-0.004}$ & ${-1.810579}$ \quad \quad ${-1.857}^{+0.079}_{-0.081}$\\
c$_{3,3}$ & \,\,\,\,\,${0.524997}$ \quad \quad \,\,\,\,\,${0.533}^{+0.018}_{-0.018}$ & \,\,\,\,\,${2.049254}$ \quad \quad \,\,\,\,\,${2.044}^{+0.011}_{-0.013}$ & \,\,\,\,\,${0.608393}$ \quad \quad \,\,\,\,\,${0.628}^{+0.034}_{-0.030}$\vspace{0.05cm}\\
\end{tabular}}
\end{ruledtabular}
\label{tab:coeffs_table_i}
\end{table*}
%-----------------------

\clearpage

%\bibliographystyle{apsrev}
\bibliography{References}

\end{document}